\RequirePackage{snapshot}
\documentclass{singlecol-new}
\pdfoutput=1
\usepackage{natbib,stfloats}
\usepackage{mathrsfs}

\usepackage[pdftex]{graphicx} \graphicspath{{./figs/}}
\DeclareGraphicsExtensions{.pdf,.jpeg,.png}

\usepackage{textcase}

\usepackage{randtext}
\catcode`\_=11\relax
\newcommand\email[1]{\_email #1\q_nil}
\def\_email#1@#2\q_nil{  \href{mailto:#1@#2}{{\randomize{#1}\emailampersat \randomize{#2}}}}

\newcommand\emailampersat{{\small@}} \catcode`\_=8\relax

\usepackage{doi}

\newcommand\hreftwo[1]{\href{#1}{#1}}

\newcommand\litem[1]{\item{\bfseries #1:}} 
\def\imagetop#1{\vtop{\null\hbox{#1}}}

\usepackage[usenames,dvipsnames]{xcolor}
\usepackage{tabularx}
\usepackage{colortbl}
\usepackage{arydshln}

\usepackage{caption}
\usepackage{subcaption}

\newcommand{\mystrut}{\vrule height 3ex depth 1.1ex width 0pt }

\usepackage{tikz}
\usetikzlibrary{calc,fit,arrows,shapes.misc}

\newcommand\markarrowtopleft[1]{    \tikz[overlay,remember picture] 
        \node (marker-#1-a) at (0,0ex) {};}
\newcommand\markarrowbottomright[1]{    \tikz[overlay,remember picture] 
        \node (marker-#1-b) at (0,0) {};	\tikz[overlay,remember picture,thick,red!100,dashed] \draw[-open triangle 45] ($(marker-#1-a.south)+(+0.07,0.06)$) -- ($(marker-#1-b.south east)+(0.09,0.06)$);}

\usepackage{fancyhdr}

\usepackage{afterpage}

\usepackage[inline]{enumitem}

\usepackage{multirow}

\def\LCF#1{\gdef\setLCF{#1}}\def\setLCF{}\def\RCF#1{\gdef\setRCF{#1}}\def\setRCF{}
\usepackage{transparent}
\newsavebox\imagebox
\newcommand*{\imagewithtext}[3][]{  \sbox\imagebox{\includegraphics[{#1}]{#2}}  \usebox\imagebox
  \llap{    \resizebox{\wd\imagebox}{\height}{      \texttransparent{0}{#3}    }  }}

\usepackage{etoolbox}
\usepackage{soul}
\sethlcolor{black}
\makeatletter
\newtoggle{@blind}
\newtoggle{noblindflag}
\togglefalse{@blind}
\iftoggle{@blind}{\def\hl#1{}\togglefalse{noblindflag}
}{\let\hl\relax
\toggletrue{noblindflag}
}

\theoremstyle{TH}{

}

\theoremstyle{THrm}{

}

\theoremstyle{THhit}{

}

\makeatletter
\def\theequation{\arabic{equation}}

\JOURNALNAME{\TEN{\it Int. J. Communication Networks and Distributed Systems,
Vol. \theVOL, No. \theISSUE, \thePUBYEAR\hfill\thepage}}
\def\BottomCatch{\vskip -10pt
\thispagestyle{empty}\begin{table}[b]\NINE\begin{tabular*}{\textwidth}{@{\extracolsep{\fill}}lcr@{}}\\[-12pt]
 & &\end{tabular*}\vskip -30pt\end{table}}
\makeatother

\begin{document}
\setcounter{page}{1}

\LRH{\hl{R. Farrahi Moghaddam et~al.}}

\RRH{A Decentralized Approach to Software-Defined Networks (SDNs)}

\LCF{\small Submitted to International Journal of Communication Networks and Distributed Systems}

\RCF{\small\begin{tabular}{c}Special Issue on Software Defined Network, Software Defined Infrastructure,\\ Network Functions Virtualization, Autonomous System \& Network Management\end{tabular}}

\VOL{x}

\ISSUE{x}

\PUBYEAR{xxxx}

\BottomCatch

\PUBYEAR{2015}

\subtitle{}

\title{A Decentralized Approach to Software-Defined Networks (SDNs)}

\authorA{\hl{Reza} \MakeTextUppercase{Farrahi Moghaddam}*}
\affA{\iftoggle{noblindflag}{Synchromedia Lab and CIRROD,\\ ETS (University of Quebec),\\ Montreal, QC, Canada H3C 1K3\\
E-mail: \email{imriss@ieee.org}\\ LinkedIn: {\small \hreftwo{https://www.linkedin.com/in/rezafm}}} *\\
* Corresponding author}
\authorB{\hl{Yves} \MakeTextUppercase{Lemieux}}
\affB{\hl{Ericsson Research Canada,\\ Montreal, QC, Canada H4P 2N2}}
\authorC{\hl{Mohamed} \MakeTextUppercase{Cheriet}}
\affC{\hl{Synchromedia Lab and CIRROD,\\ ETS (University of Quebec),\\ Montreal, QC, Canada H3C 1K3}}

\begin{abstract}
Redistribution of the intelligence and management in the software defined networks (SDNs) is a potential approach to address the bottlenecks of scalability and integrity of these networks. We propose to revisit the routing concept based on the notion of regions. Using basic and consistent definition of regions, a region-based packet routing called SmartRegion Routing is presented. The flexibility of regions in terms of naming and addressing is then leveraged in the form of a region stack among other features placed in the associated packet header. In this way, most of complexity and dynamicity of a network is absorbed, and therefore highly fast and simplified routing at the inter-region level along with semi-autonomous intra-region routing will be feasible. In addition, multipath planning can be naturally realized at both inter and intra levels. A basic form of SmartRegion routing mechanism is provided. Simplicity, scalability, and manageability of the proposed approach would also bring future potentials to reduce energy consumption and environmental footprint associated to the SDNs. Finally, various applications, such as enabling seamless broadband access, providing beyond IP addressing mechanisms, and also address-equivalent naming mechanisms, are considered and discussed. 
\end{abstract}

\KEYWORD{Packet Switching; Decentralized Routing; Region-based Routing.}

\iftoggle{noblindflag}{\begin{bio}
Reza \MakeTextUppercase{Farrahi Moghaddam} received his B.Sc. degree in Electrical Engineering and his Ph.D. degree in Physics from the Shahid Bahonar University of Kerman, Iran, in 1995 and 2003, respectively. He has been a Postdoctoral Research Fellow and a Research Associate with the Synchromedia Laboratory for Multimedia Communication in Telepresence, \'{E}cole de technologie sup\'{e}rieure (University of Quebec) in Montreal (QC), Canada since 2007 and 2012, respectively. Reza has published more than 50 technical papers. His research interests include sustainability, behaviour analysis, green ICT, green economy, visual perception, and optimization. He is a member of the IEEE.\vs{9}

\noindent Yves \MakeTextUppercase{Lemieux} joined Ericsson in 1994 and currently holds the position of Research Engineer in Cloud Technology group in Ericsson Canada Inc. Yves has a number of patents and publications to his credit in the fields of Cellular System Synchronization Selection, Network Resiliency, LTE Core Network Congestion Control, among others. His main interests are now vested in 3GPP based End-to-End QoS and also Virtualization for Cloud Computing.
Prior to working at Ericsson, Yves was a Systems Design Engineer at AT\&T Canada and a Radio/Fiber Manager at Rogers Wireless.
Yves received his bachelor degree in Electrical Engineering from the University of Sherbrooke in December 1981, and a Masters Degree in Computer Engineering from Polytechnique in Canada, in June 2005.\vs{8}

\noindent Mohamed \MakeTextUppercase{Cheriet} received M.Sc. and Ph.D. degrees in Computer Science from the University of Pierre et Marie Curie (Paris VI) in 1985 and 1988 respectively. Since 1992, he has been a professor in the Automation Engineering department at the \'Ecole de technologie sup\'erieure (University of Quebec), Montreal, and was appointed full Professor there in 1998. Mohamed is the founder and director of Synchromedia which targets multimedia communication in telepresence applications. 
Mohamed's research has acquired extensive experience in cloud computing and network virtualization. In addition, Mohamed is an expert in Computational Intelligence, Pattern Recognition, Mathematical Modeling for Image Processing, Cognitive and Machine Learning approaches and Perception. Mohamed has published more than 350 technical papers in the field.  He serves on the editorial boards of several renowned journals and international conferences. Mohamed holds Tier 1 Canada Research Chair on ``Sustainable Smart echo-Cloud,'' and the Canada Foundation for Innovation funding to build the first Smart Residence at ETS campus, to serve as a scale model for Smart Montreal District. He is the recipient of the Queen Elizabeth II Diamond Jubilee Medal. He is a senior member of the IEEE, and the founder and former Chair of the IEEE Montreal Chapter of Computational Intelligent Systems (CIS).
\end{bio}
}

\maketitle

\section{Introduction}
\label{sec_introduction}
The SDN concept has greatly helped to separate the actual `less-smart' data handling elements of a network from the rest by decoupling the control plane from the data plane \citep{Kreutz2014,Jacob2014,Alvizu2014,Feamster2014}. Among various realizations of SDNs, OpenFlow-based approaches have received a great interest thanks to their minimal footprint on the switch logic while providing a straightforward way to modify and change the forwarding tables on-the-fly using OpenFlow controllers \cite{McKeown2008,Bianchi2014}, \cite{Bianchi2014a}.

\begin{figure}[!thb]
\centering
\setlength{\tabcolsep}{2pt}
\begin{tabular}{@{}cc@{}}
(a) & \fbox{\includegraphics[width=4.5in]{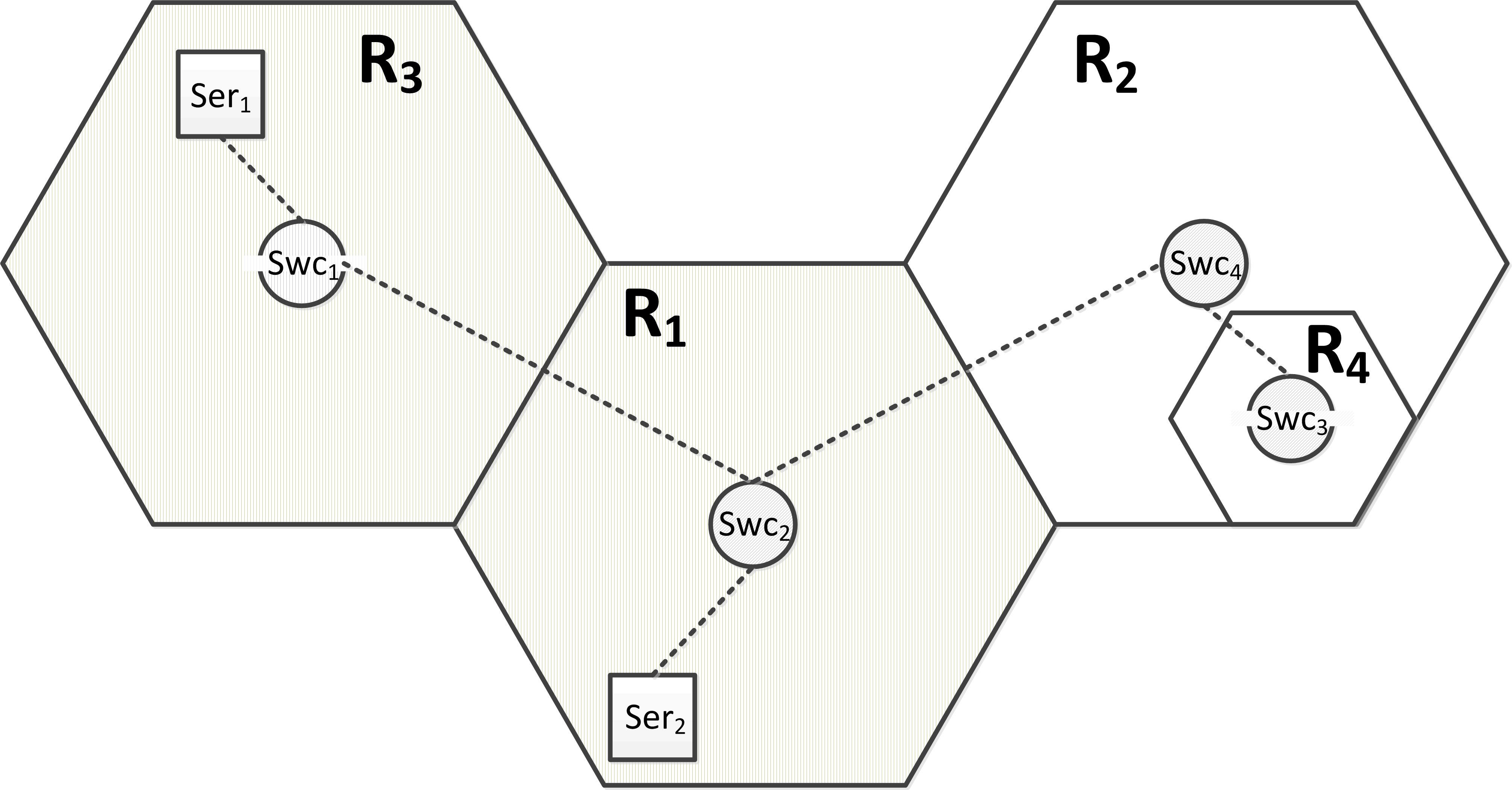}} \\
(b) & \fbox{\includegraphics[width=4.5in]{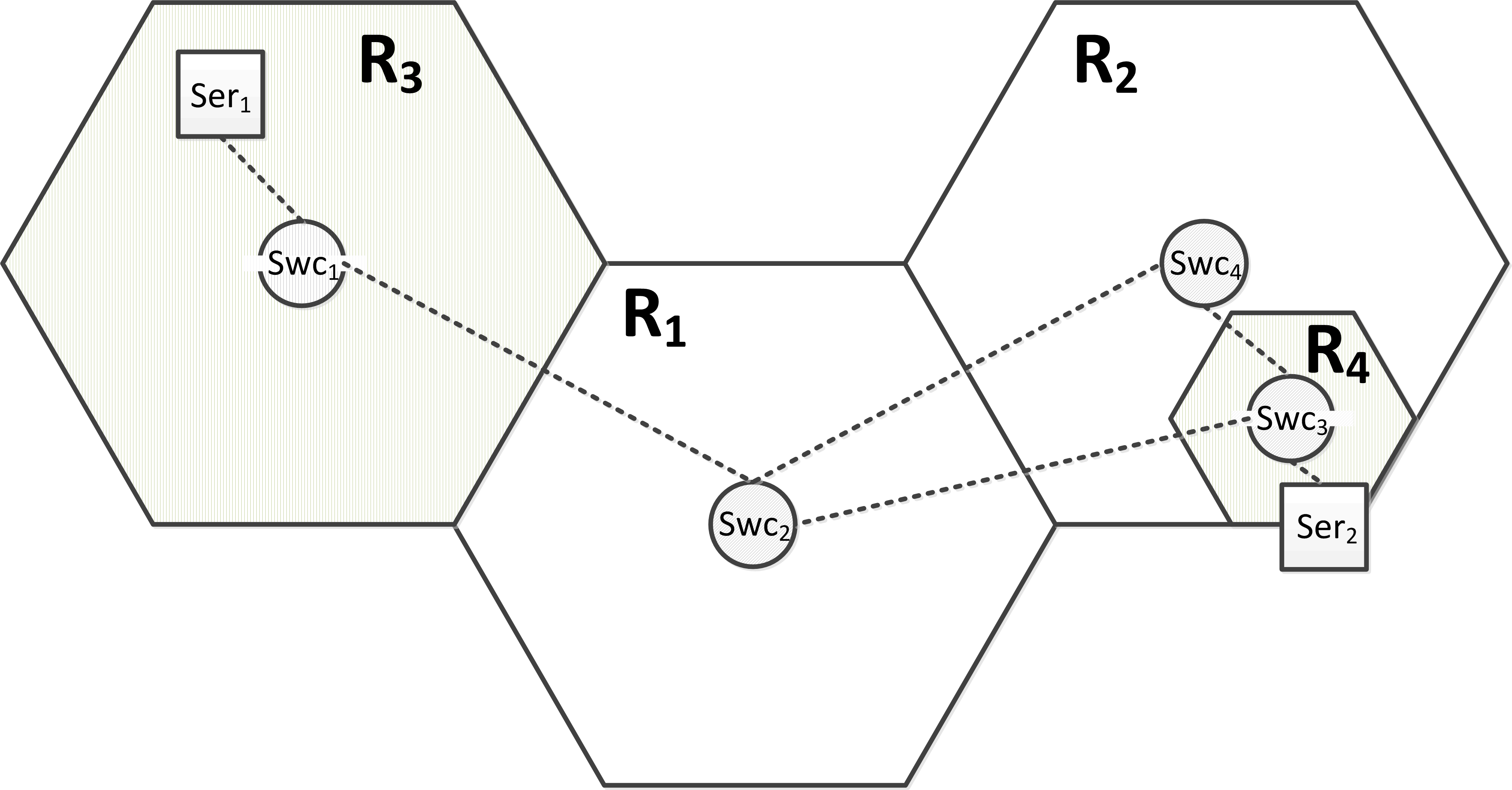}} \\
\end{tabular}
\caption{A typical example of mobility in a network. a) The network topology before a change. b) The topology after displacing node $\textbf{Ser}_2$.}
\label{fig_VN_Triangle_1}
\end{figure}

The main problem with controller-based SDNs is their high-level of dependency of the actions under the controller's command. It has been observed that this could highly impact the performance in terms of both adaptability to changes and also scalability \cite{Alvizu2014,Farrahi2015}. 
Changes and mobility are unavoidable in networks. A clear example is the case of the broadband wireless and Telecom networks. The long term evolution-advanced and its enhancements, usually coined 5G, targets providing semi-symmetric 1 Gbps broadband access to individual mobile users \cite{Eichinger2014,NTTDOCOMOInc2014,Ericsson2015,Ericsson2015a,4GAmericas2014,NGMN2015}, \cite{Cho2014}. Figure \ref{fig_VN_Triangle_1} shows a change in the configuration of a network because of mobility of a node. Dynamic changes in the network topology are not limited only to the mobile networks. In any virtual or private network, a virtual machine node or a container could be simply migrated to or be restarted at another persistent location. Although the lowest level of the stack of multiple overlapping virtual networks would be the same shared persistent physical layer, the network functions of a virtual network would require to be adapted to any change. Traditional IP-based approaches would be inefficient in all these cases because there is no relation between the node IDs, i.e., the IPs, and the actual `location' of the nodes. This has been the motivation for decoupling the actual node IDs from their location IDs  \cite{ONeill2014} (see section \ref{sec_related_work} for more discussion). In another approach, in this work, this decoupling is achieved using a region-based representation of the networks and also enabling packets to carry routing directions (see also \ref{sec_suppl_Decentralized_Routing_Intelligence_SmartRegion}).

In this SmartRegion routing, `regions' are used to determine the location of a node. In particular, the immediate region hosting the destination node of a flow is considered as the destination location. In case of big networks, a hierarchy of embedding regions is used. Also, the regions are considered fuzzy, i.e., they could overlap on some of the nodes regardless of their level in the hierarchy. A region stack is considered in order to allow a flexible way to augment and to precise the path to a destination. If the region stack is fully described from the source to the destination, it could represent a preferred or `trusted' path selected by the flow participants. In particular, the region stack provides some sort of high level control implicitly delivered from the packet side not from the routing tables. That means that there will be less requirement to globally and dynamically update and set the tables, which has been a blocker in terms of scalability. The main difference between the proposed approach and that of   \cite{ONeill2014} is that in our case we use less granular concept of regions to express the location of a node. This is a big advantage for our approach in terms of scale and simplicity. Also, multipath routing could be achieved without a central planning; the packet is needed to be delivered to a specific `region' not to a specific `switch.' The handling of the packet within an intermediate region is performed by the collective actions of all switches of that region (or possibly a regional controller), which could deliberately execute a multipath intra-region routing. In addition, multipath routing at the inter-region level is enabled and recommended when a full-path region stack is not specified (see \ref{sec_suppl_Multipath_Routing_SmartRegion}). The full description of the SmartRegion approach is provided in the following sections. It is worth mentioning that advanced analytic approaches to routing, such as that described in this work, would be implementable and feasible considering increasing level of programmability in the networking devices including i) development of smart ASIC-based switches \cite{Bosshart2013}, \cite{Bosshart2013a}, ii) adaptation of white box switches, and iii) direct access between CPU/GPU and I/O.

In terms of constraints of this work, we assume a static network in that sense that the node IDs (NIDs) are unique and would not change along the time. This does not rule out having a WAN or geographically-distributed network. It is also assumed that a name (or identity) provider (IdP) service \cite{Kreutz2014a} is presented that handles the uniqueness of NIDs.  Cases where networks are dynamically merged/separated, and ultimately the case of the Internet would be considered in another work (see also \ref{sec_suppl_Dynamic_Actions_SmartRegion}).
Also, we postpone analyzing the impact of the asymmetrical nature of many links, especially in the access regions, to another work. 
\iftoggle{noblindflag}{It is also worth mentioning a shorten version of this work was presented in \cite{Farrahi2015}.
}

In addition to packet switching, other forms of switching such as circuit switching and burst switching are in practice especially in optical fabrics and inter-data center networks \cite{Alvizu2014,Vadrevu2013,Sadasivarao2013}. Our approach could be generalized to also cover them. We leave this generalization to a future work.

The paper is organized as follows. First, a list of challenges in front of SDNs from various perspectives is provided in Section \ref{sec_Challenges}.
In Section \ref{sec_Basic_Definitions}, the basic concepts used in the rest of the paper are discussed.
The details of the SmartRegion routing is provided in Section \ref{sec_SmartRegion_Routing}. Firstly, the header part of SmartRegions is presented in the form of four super fields in Section \ref{sec_SmartRegion_Header}. Then, the routing behavior in both static and transition modes are discussed in Sections. In particular, illustrative examples are provided to show the flexibility and also the details of the proposed approach. Some applications are discussed in  section \ref{sec_Discussions} with a focus on new applications using the concept of regions. Then, the conclusions and some future prospects are provided in Section \ref{sec_conclusion}. Finally, an alternative variation of the SmartRegion approach's definition of regions is presented in the Appendix.

\section{Related Work}
\label{sec_related_work}
Because of the limited space, we briefly provide the related work.
In terms of locality in routing, \cite{Moshref2014} could be mentioned in which a `dynamic' approach was used to update the rules of the switches using the `local' data. In \cite{Bianchi2014}, \cite{Bianchi2014a}, a platform-agnostic programming called OpenState was proposed to rescind the reliance on external controllers. In another approach, nodes IPs were separated from their `location identities` in order to make policies and packet switching aware of locations especially in SDNs where the IPs do not benefit from the same level of aggregability as in the traditional networks. Packets that carry instructions were considered in \cite{Jeyakumar2013,Jeyakumar2014}. In another approach, forwarding using parse-and-match was used toward protocol-independent packet processing \cite{Bosshart2013}. 
In \cite{Schwartz1999}, smart packets were used to transfer programs, data, and messages. In contrast, the proposed SmartRegion here focuses on enriching packet switching toward simplicity and scalability. In another approach to decentralization of controller functions, multiple controllers sharing a common, high-performance central data store was proposed in \cite{Botelho2013}, and called SMaRtLight. In addition, in \cite{Gelenbe2014}, awareness of QoS requirements and interests were considered in an approach to cognitive packet network.

Here some work on routing approaches at the Internet level, including inter-domain routing, is listed. 
Recently, a domain-level OpenFlow controller was demonstrated in which each domain has its own controller and there was a global controller for all domains \cite{Brown2014}.
In terms of routing at the domain level, various protocols are used \cite{Sobrinho2012,Sobrinho2014}: i) Border Gateway Protocol (BGP) \cite{Rekhter2006,Elmokashfi2010}, 
ii)  Path Computation Element (PCE)-based computation \cite{King2012},
and iii) Backward Recursive PCE-based Computation (RBPC) \cite{Vasseur2009,Xu2014}. These solutions are highly adapted to the Internet architecture, and implicitly require full autonomy of every domain. Considering the definition of a domain which is a collection of network elements within a common sphere of address management or path computational responsibility \cite{Farrel2006}, such as an Interior Gateway Protocol (IGP) area or an Autonomous System (AS), adapting these protocols to other networks would require special considerations, and might not be an optimal solution.
In addition, these approaches are more suitable for circuit switching because a particular planned path can be reused for a large amount of transfer. 
In contrast, the SmartRegion routing is based on region decomposition which does not require full autonomy or full dedication, while it provides possibility to do so.

\section{Challenges}
\label{sec_Challenges}
In this section, the challenges in front of the future virtualized networks are briefly listed.

\subsection{Physical}
Although our focus is on the SDN in this work, IP-based networking (both IPv4 and IPv6 versions) has a big influence on the virtualized networks. Therefore, the limitations of the IP networks have been transferred to the SDN in some degree. Various mechanisms in order to relax the limitation of structured nature of the IP addressing have been practiced including Classless Inter-Domain Routing (CIDR), Variable-Length Subnet Masking (VLSM) \cite{Narten2011,Fuller2006} and Autonomous System (AS), and Interior Gateway Protocol (IGP) \cite{Farrel2006,Vasseur2009,Dhody2014}. However, the aggregation-oriented nature of IP networks is gradually becoming less efficient mainly because of mobility of the nodes. In particular, providing high quality of experience for broadband access on fast moving means of transport, such as fast trains \cite{Ecotrain2011}, is imposing major challenges especially in cases of intermittent access along the route. that Moreover, the numerical aggregation of IPs is be definition less adaptive compared to possible set-based aggregation approaches. In addition, requiring end-to-end addressable (IP) networks while the penetration level Point/Home and Person levels because the networks would be less aggregatable and more meshed \cite{Kawai2014}.

\subsection{Virtual}
At the virtualized layer of  SDNs, the positive impact of aggregating the virtual IP would become minimal, and even it can be argued that it would degrade the performance even compared to no-aggregation practice. In other words, even if the underlying physical layer is aggregatable, there is no meaningful relation between the virtual IPs and the associated physical ones in a `true' virtualized layer, and therefore any attempt to aggregate using the virtual IPs would be a random and worthless action. Moreover, in the move toward Network Function Virtualization (NFV), the elements of the virtual network would be `non-local' and mobile. Therefore, even if the virtual network is in a IP aggregatable state at a moment, the state would change a short period of time because of change in the associated service request volume and therefore dynamicity in the chaining of the functions.

\subsection{Central}
The main challenges in front of SDNs root in the current centralized approaches used in practice. Although these approaches are efficient at small and medium scales, they suffer from considerable amount of planning overhead and also extra latency related to requirement to reach furthest switches when the scale of network increases. Especially in large-scale highly-dynamic networks, the central controller would lag in a delayed state picture of network would degrade the performance. Although multi-controller and multi-domain alternatives have been considered \cite{}, it seems that a scale-friendly approach would be more toward `management' of `local', non-autonomous sub-networks that `control' of `whole' network. This is one of our motivations to propose a region-based approach in this work.

\section{Regions, Region Decomposition, and Region Maps}
\label{sec_Basic_Definitions}
The regions of the proposed SmartRegion approach are defined with a specific purpose of having a high degree of disaggregation. We call it the DISAGG variation. However, it is worth mentioning that this is not the only possibility. For example, in \ref{sec_Alternative_Definition_REACH}, an alternative approach to definition of regions is provided that emphasizes on the reachability of nodes. 

A main advantage of a region-based approach is its ability to describe the connectivity of a network in both horizontal and vertical dimensions by definition. The horizontal dimension could easily help in absorbing the `dynamicity' of the network and containing it within small regions. The hybridization of horizontal and vertical connectivity would require and at the same time provide means for name-based addressing, which goes beyond the capabilities of number-based approaches (we will discuss this point in more details in Sections \ref{sec_app_in_addressing} and \ref{sec_app_in_naming}). In addition, multipath routing would be implicitly included in an extended form, which covers spatial and spectrum dimensions, and would not require a ubiquitous, central network governance down to the intra-region level, while the network would be dependently manageable at both inter- and intra-region levels avoiding unmanaged flooding.

A region-based approach, with its capability to differentiate between inter-region and intra-region network governance, would enable federated and confederated network governance, which automatically would result in  seamless scalability with linear overhead and without requiring to propagate the inter-level changes to intra-level controllers and vice versa. This would lead to quick convergence because in most cases only neighboring regions and inter-level service would be required to be Informed.

We start will basic definitions and finally the Region Decomposition (RD), Region Map (RM), and Region Path (RP) concepts are presented:

\begin{figure}[!t]
\centering
\begin{tabular}{@{}c@{}@{}c@{}}
\imagetop{\fbox{\includegraphics[width=2.6in]{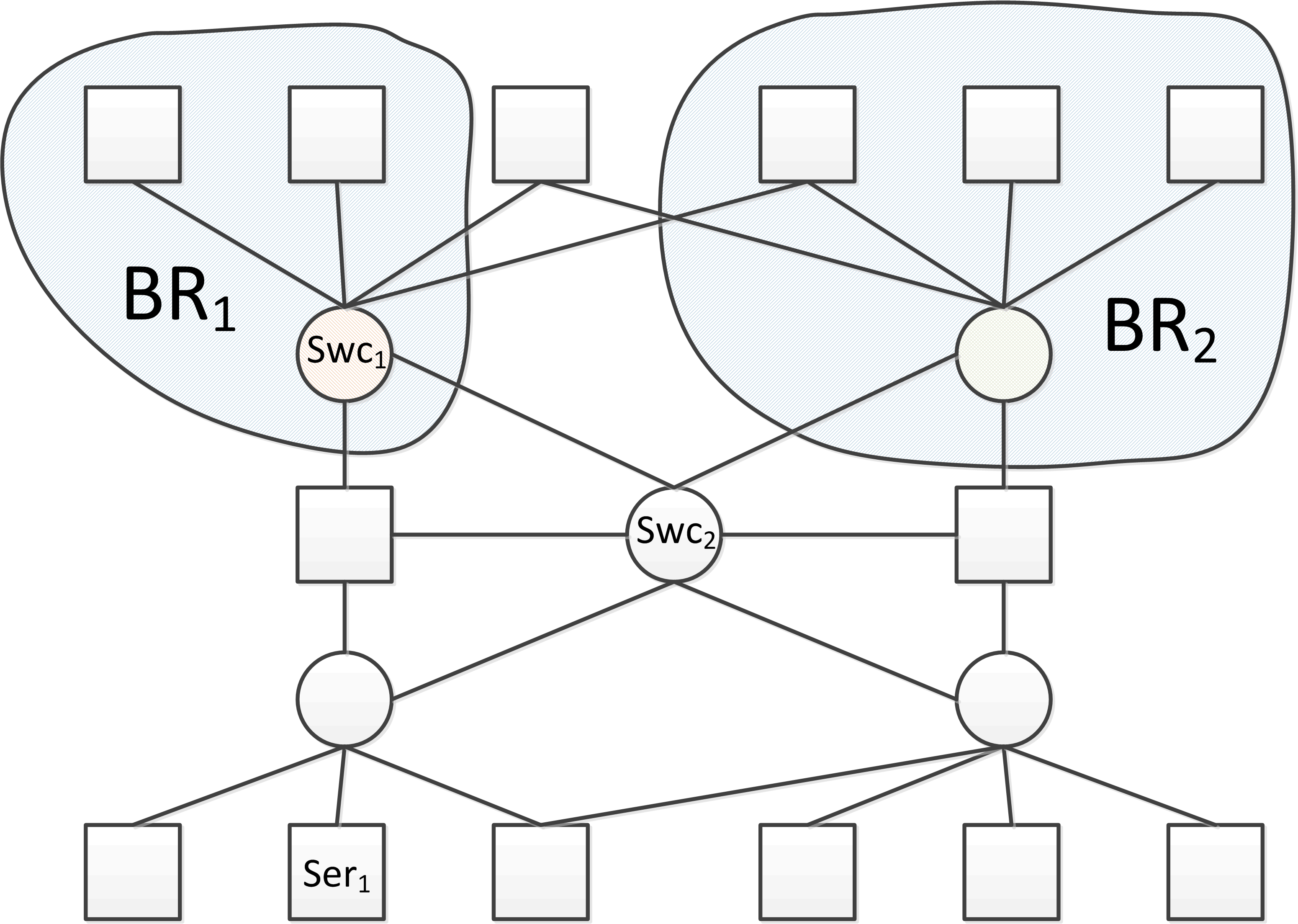}}} &
\imagetop{\fbox{\includegraphics[width=2.6in]{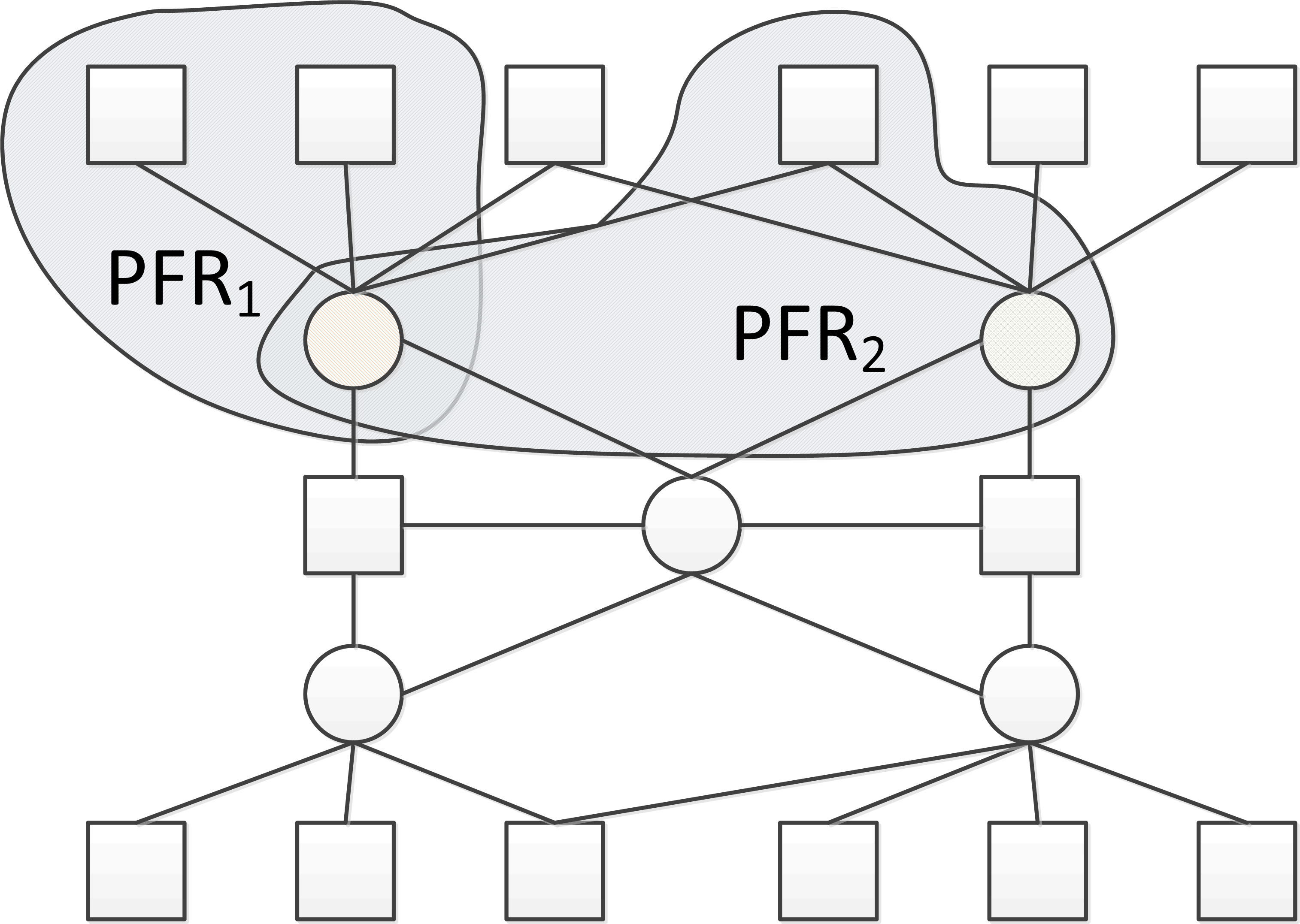}}} \\
(a) & (b) \\
\imagetop{\fbox{\includegraphics[height=1.95in]{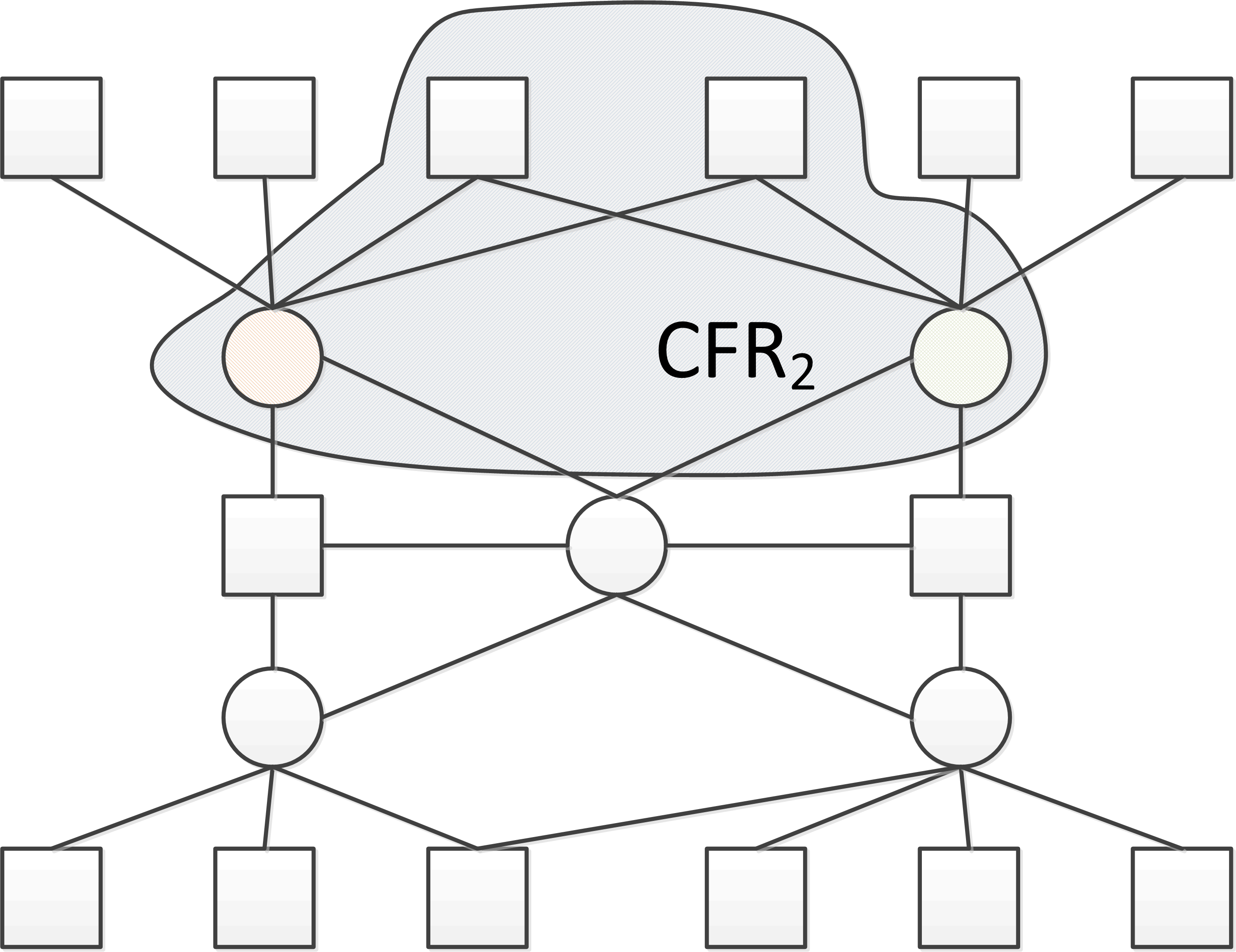}}} &
\imagetop{\fbox{\includegraphics[height=1.95in]{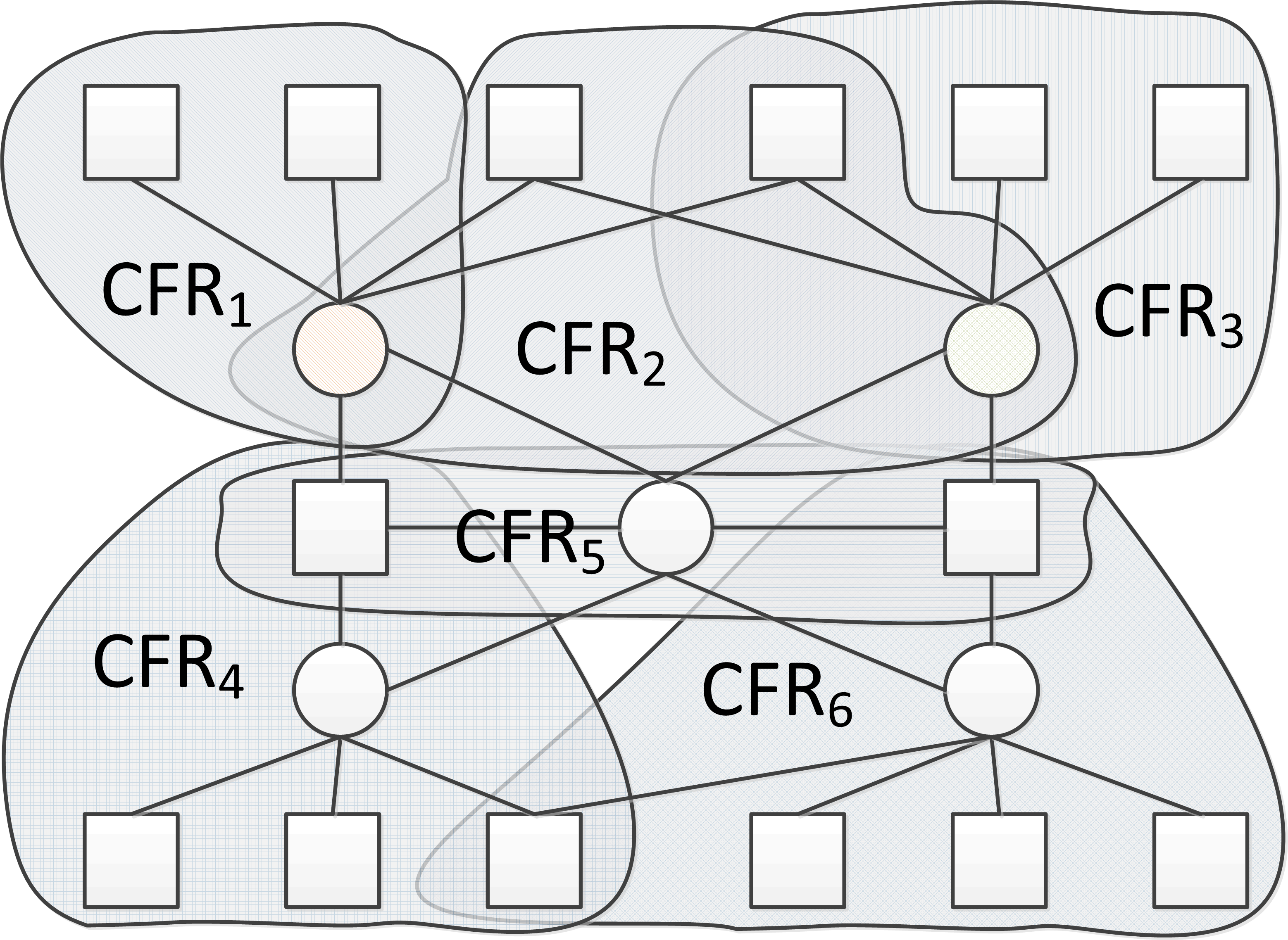}}} \\
(c) & (d) 
\end{tabular}
\caption{
A typical example of how the Basic Regions (BRs), Partial-Full Regions (PFRs), and Complete-Full Regions (CFRs) is provided in (a), (b), and (c) respectively. A possible Region Decomposition (RD) is shown in (d).
}
\label{fig_PFR_CFR_Regions_1}
\end{figure}

\subsection{Node} A computer unit whose main function is not networking. However, a node can also participate in packet routing actions of other nodes,for example, in a BCube-like topology \citep{Farrahi2014g}. In Figure \ref{fig_PFR_CFR_Regions_1}, nodes are shown as squares. To be specific, in this case, we assume a form of collocation of the node and a switch in the same boundary (see Figure \ref{fig_Regions_BCube_1}(a)). From the NFV perspective, every virtualized network function would collocate with other virtual resources in the boundary of the hosting resource. In this work, we limit the scope only to switches as defined below.

\subsection{Switch} A computer unit whose main function is routing among other network functions. A switch does not initiate a flow (with possible exceptions mentioned in \ref{sec_suppl_Dynamic_Actions_SmartRegion}). We may also occasionally use the term node for switches. In Figure \ref{fig_PFR_CFR_Regions_1}, nodes are shown as circles. 

\subsection{Network} A network is an undirected graph of nodes and switches as its vertices. The links between the nodes and switches are considered as the edges of the associated network.

\subsection{Basic Region (BR)} A basic region is a connected subgraph of a network that inherited all applicable edges. Figure \ref{fig_PFR_CFR_Regions_1}(a) shows examples of BRs.

\subsection{Partial-Full Region (PFR)} A partial-full region is a basic region in which every switch is connected to `all' nodes. Figure \ref{fig_PFR_CFR_Regions_1}(b) shows examples of PFRs.

\begin{figure}[!t]
\centering
\setlength{\tabcolsep}{2pt}
\begin{tabular}{@{}cc@{}}
\imagetop{\fbox{\includegraphics[height=1.86in]{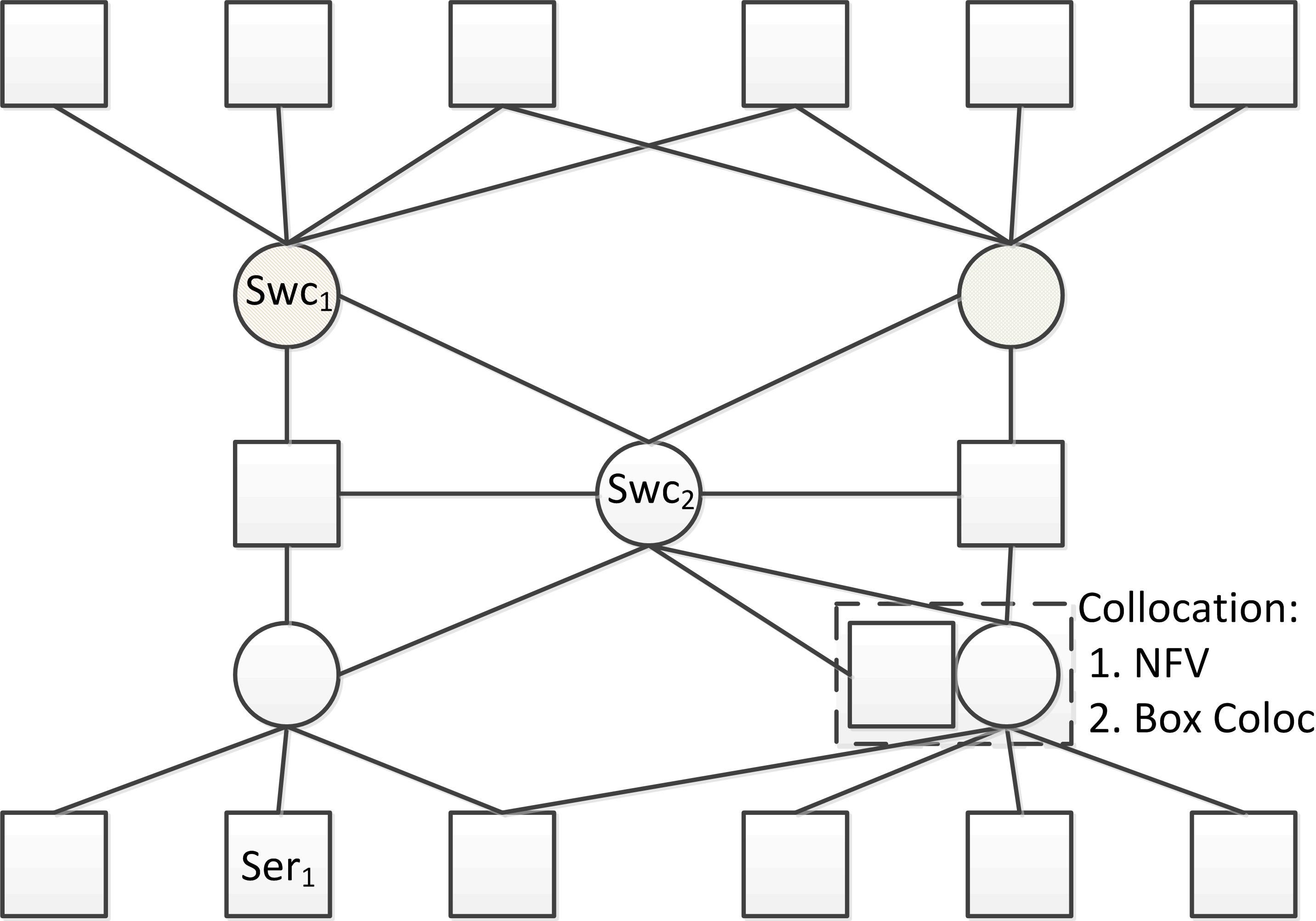}}} &
\imagetop{\fbox{\includegraphics[height=1.86in]{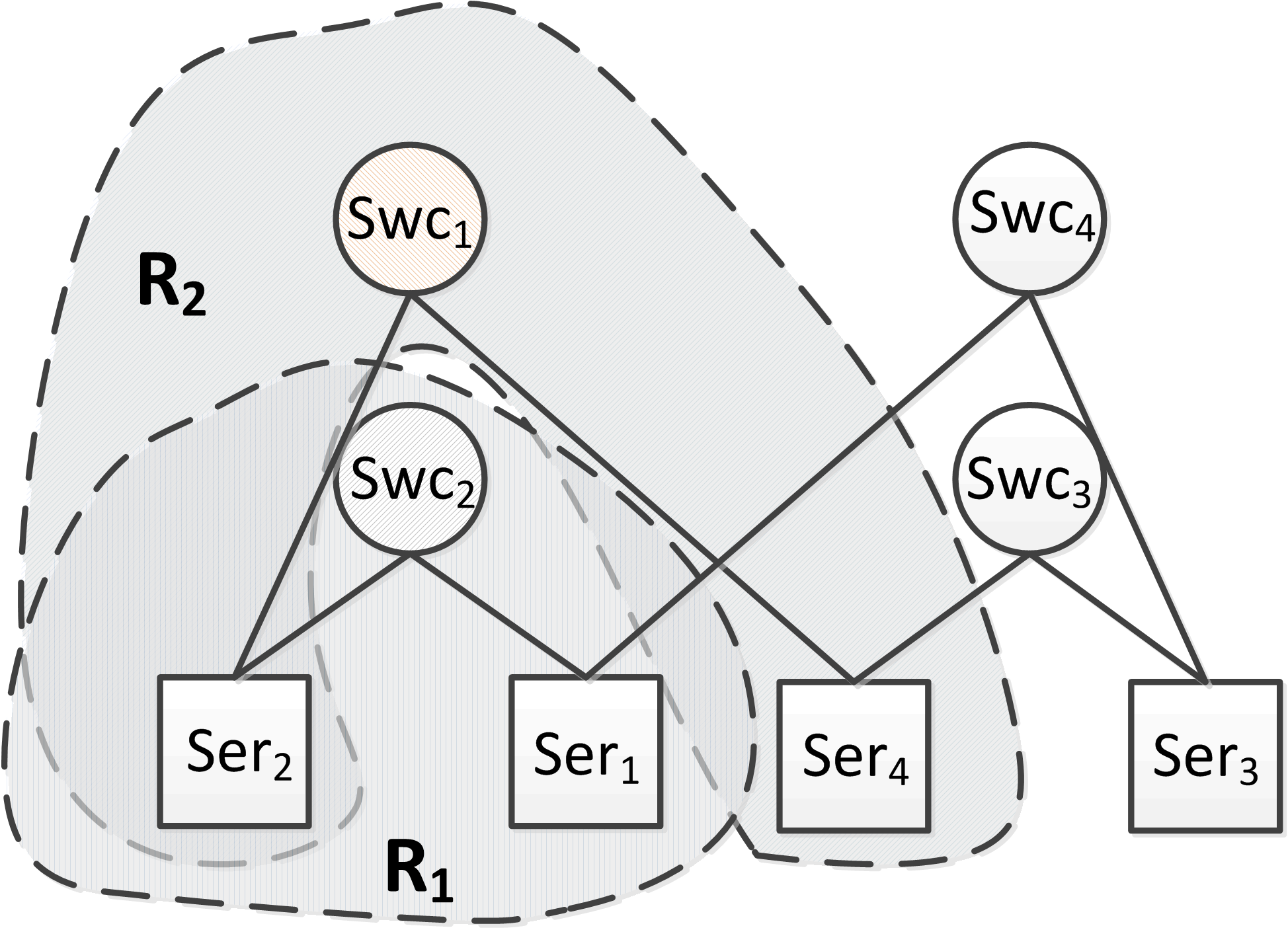}}} \\
(a) & (b)
\end{tabular}
\caption{
a) An example of possible (either NFV or physical-box) collocation in a network.
b) An examples of two overlapping (complete) full regions in a BCube-like access sub-network.
}
\label{fig_Regions_BCube_1}
\end{figure}

\subsection{Complete-Full Region (CFR)} A PFR which is not a subgraph of another, bigger PFR with the `same' set of switches. Figure \ref{fig_PFR_CFR_Regions_1}(b) shows examples of CFRs. Also, Figure \ref{fig_Regions_BCube_1}(b) provides examples of CFRs in a BCube-like access sub-network. We may use the term {\em full region} in place of the {\em complete full region} term from hereon.

\begin{figure}[!t]
\centering
\setlength{\tabcolsep}{2pt}
\begin{tabular}{@{}cc@{}}
\imagetop{\fbox{\includegraphics[width=2.6in]{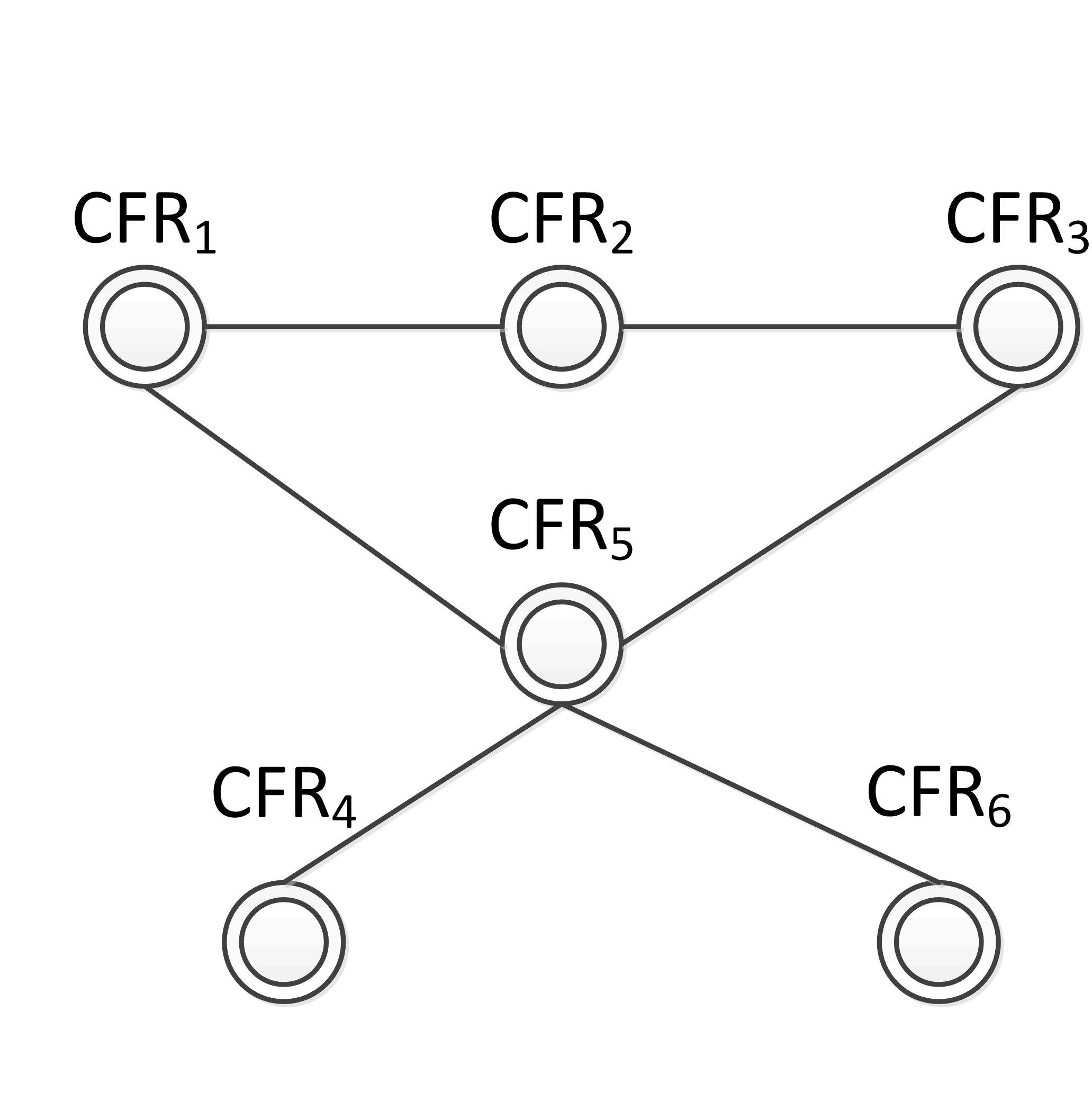}}} &
\imagetop{\fbox{\includegraphics[width=2.6in]{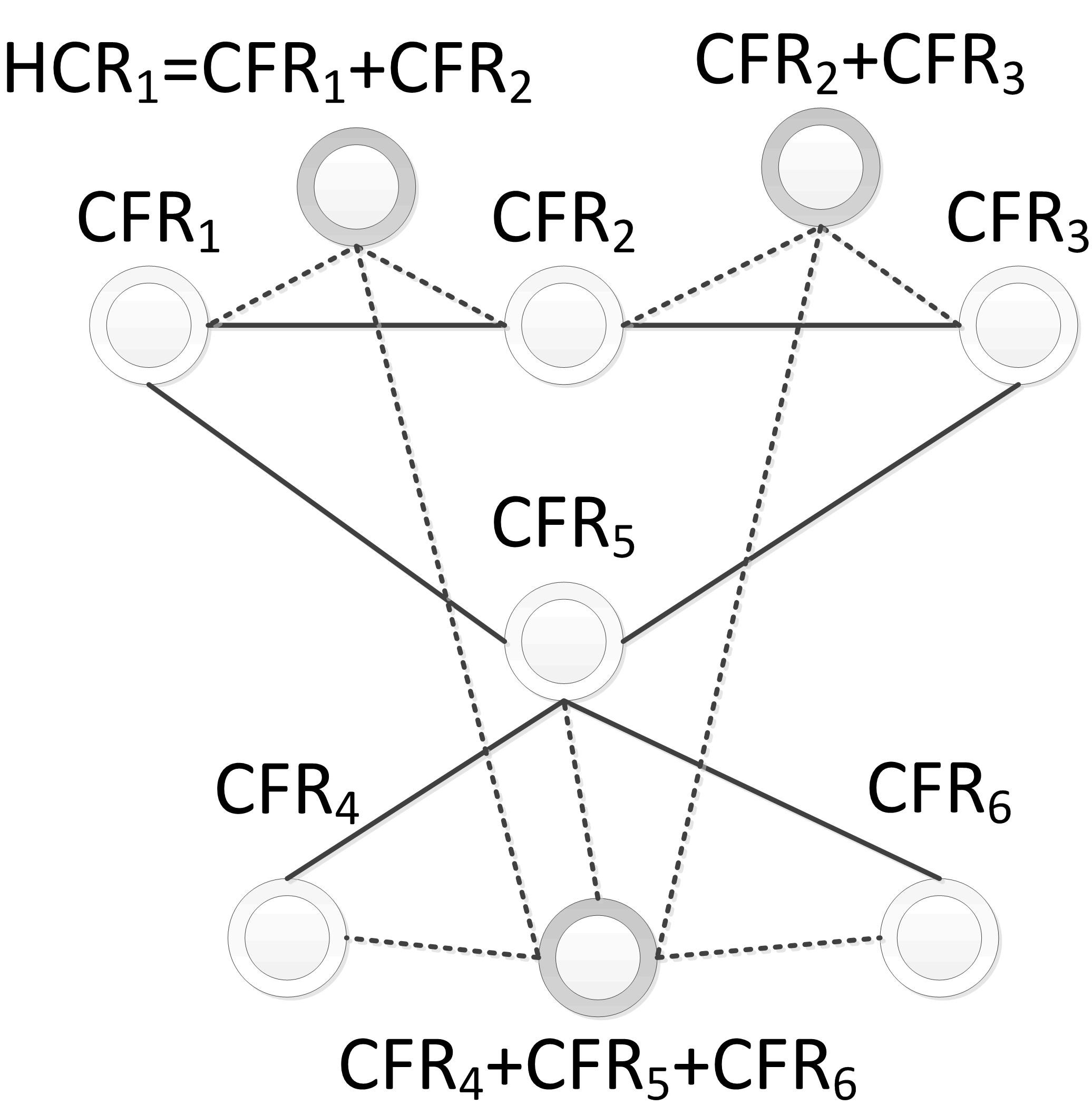}}} \\
(a) & (b)
\end{tabular}
\caption{
a) An example of RM associated to the RD in Figure \ref{fig_PFR_CFR_Regions_1}(d).
b) An extension of (a) when some HCRs are included.
}
\label{fig_Region_Map_1}
\end{figure}

\subsection{Region Decomposition (RD)} A subset of the set of all CFRs augmented with some {\em High-level} Complex Regions (HCRs) recursively generated by combining full regions or high-level regions. 
The hierarchical nature of HCRs provides a direct set-based aggregation capability when number-based aggregations, such as that of IP, fail.
In future, a weak form definition of region decomposition will be considered in which PFRs are also allowed. However, here we limit a region decomposition only to the complete-full regions.

\subsection{Region Map (RM)} An undirected graph in which vertices are `regions' of a region decomposition.

\subsection{Region Path (RP)} A region path from a sender node $N_S$ to a receiver node $N_R$ is represented by a resolved sequence of regions selected from a region decomposition. Within each region in this sequence, the region path is later on resolved to the actual switch-level [multi-]path by the (virtual) controller of that region.

\section{SmartRegion Routing}
\label{sec_SmartRegion_Routing}

As mentioned before, in the proposed approach, we use `regions.' In particular, the last region to the destination node is used as the guideline to the destination location. In case of big networks, a hierarchy of embedding regions may be used. Also, our regions are fuzzy in that sense that they can overlap on some nodes even if they are at the same level. We also allow chaining of the regions in cases that the sender-receiver have an agreement on a `trusted' path of regions to pass the packets through. Chaining can be also used to implement some sort of high level controlling enforced from the packet side not from the mapping tables, and therefore avoid necessity to update the tables dynamically, or guess them at the beginning with high chance of error. The main difference between ours and  \cite{ONeill2014} is that in our case a region is defined as a collection of nodes and switches with a condition that all nodes are connected to all switches. 
In addition, it can be argued that IP is not a good approach to assign a unique `name' to a node; historically an IP has been equivalent to geographical location of a node, and therefore when a node for example with Canadian IP is moved to a USA-based data center, there is a high chance of an implicit violation of an `unwritten' service agreements that were based on the assignment of a Canadian IP to that node at first.

\subsection{SmartRegion Header}
\label{sec_SmartRegion_Header}
In the proposed SmartRegion routing, the header of each packet would be composed of four SuperFields (SFs). Each SF would be itself composed of a list of fields at predefined positions or arranged in the form of a stack. Namely, we consider the following SuperFields: i) Region Stack SF, ii) IDs SF, iii) QoS Smart SF, and iv) {Backward Region Stack (BRS) SF (Table \ref{tab_SmartRegion_Header_1}; see also \ref{sec_suppl_Basic_Header_Reservation_Syntax_SmartRegion}). The full description of each SF is provided below. In short, the Region Stack SF is the core of the proposed routing approach, and it provides a flexible and at the same time simple non-central approach to packet routing.

\begin{table}
\centering
\scriptsize
\setlength{\tabcolsep}{2pt}
\begin{tabular}{@{}>{\mystrut}cccc@{}}
\imagetop{\begin{tabular}{||l|l||}\hline\hline
\multicolumn{2}{||c||}{Region Stack SF} \\\hline
Index & Region RID \\\hline\hline
(2) & \begin{tabular}{@{}l}Ephemeral\\ Single-hop FID\end{tabular} \\\hline
(1) & Intra-region FID \\\hline
0 & $R_\alpha$ \\\hline
-1 & $R_\gamma$ \\\hline
-2 &  $R_\eta$ \\\hline
-3 &  $R_R$ \\\hline\hline
\end{tabular}
} &

\imagetop{\begin{tabular}{||l|l||}\hline\hline
\multicolumn{2}{||c||}{IDs SF}\\\hline
Field & Description \\\hline\hline
(Packet PID) & Packet ID \\\hline
(Flow FID) & Flow ID \\\hline
Sender NID & \begin{tabular}{@{}l}Node ID of\\ the sender\end{tabular} \\\hline
Receiver NID &  \begin{tabular}{@{}l}Node ID of\\ the receiver\end{tabular} \\\hline\hline
\end{tabular}
} &

\imagetop{\begin{tabular}{||l|l||}\hline\hline
\multicolumn{2}{||c||}{QoS Smart SF}\\\hline
QoS Metrics & Value \\\hline\hline
\begin{tabular}{@{}l}(Single-hop\\ Latency)\end{tabular} &  \\\hline
(Path Latency) &  \\\hline
(Single-hop Loss) &  \\\hline
(Path Loss) &  \\\hline
(Fission Rate) & 1\ (Default), \\
& $>$1 (RedunCast)\\\hline\hline
\end{tabular} 
} &

\imagetop{\begin{tabular}{||l|l||}\hline\hline
\multicolumn{2}{||c||}{\begin{tabular}{@{}l}Backward Region Stack\\ (BRS) SF\end{tabular}}\\\hline
Index & Region RID \\\hline\hline
(1) & $R_\xi$ \\\hline
(2) & $R_\phi$ \\\hline
(3) &  $R_\theta$ \\\hline
(4) &  $R_S$\\\hline\hline
\end{tabular} 
}
\end{tabular} 
\caption{A typical header of a SmartRegion. The items marked with parentheses are case based and are not obligatorily.}
\label{tab_SmartRegion_Header_1}
\end{table}

\begin{LaTeXenumerate}
\litem{Region Stack SuperField} This SF is arranged in the form of a stack from the closest/highest region to farthest/lowest region relative to the source node. The regions' ID ( RIDs) are not required to be unique, as long as they can be resolved using their higher level regions. 
The Region Stack could be preceded by the Intra-region FID in case the current region has assigned one (or more) ephemeral intra-region flows to the associated flow. In this case, the switches could simply skip processing of the rest of the stack and execute the formula of that (or those) intra-region flows. In addition, for low-bandwidth and high-loss hops, the FID of an ephemeral single-hop (or multi-hop) flow that is used to transfer the packet across that hop(s) could be inserted at the beginning of the stack.
\litem{IDs SuperField} This SF provides the unique NID of the sender and receiver nodes in addition to possible flow FID and packet PID. If the receiver NID is not set, the flow would be forward to all nodes of the destination region, and this provides an implicit and decentralized way for multicast routing.
\litem{QoS Smart SuperField} This SF bring smartness in the form of a set of rules (such as QoS rules) that are used to choose among various region-paths that a switch suggests. For each suggestion there is a QoS measures (such as latency to the receiver, or latency to the next switch) and at the same time some `costs' or fees that the sender/receiver will bear if that path is chosen (usually it is the sender that should take the cost because the packet's route (outside the last-mile access area) should be smartly set without disturbing the receiver. 
\litem{Backward Region Stack (BRS) SuperField (optional)} This optional stack saves the regions that the packet has been traversed up to the current switch. This information would be the result of curtsy of the traversed switches, and any of those switches may delete the data in this field. However, the side effect for such switches would be that their associated regions (and also the shadowed regions, i.e., the regions that have those switches as some sort of gateway for some other regions) would be blackholed or sent for second screening 
\iftoggle{noblindflag}{\cite{Farrahi2014e} 
}
because they cannot be discriminated from the attackers (see section \ref{sec_Advanced_Response_DDoS}). 
\end{LaTeXenumerate}

In addition, it could be mentioned that the Region Stack SF could be inserted at the beginning of a packet by a switch in case a participating sender node is not able to generate SmartRegions. In such cases, the other SFs will be excluded, and the Region Stack SF will be followed by the actual header of the received packet. Practically, the information of this header is used by the first receiving switch to identify the regions and generate the associated Region Stack SF. This patch-like workaround could enable SmartRegion-aware networks operate with incompatible nodes and transfer their flows. However, the flows from such nodes would not benefit from QoS-aware and other advantages of the proposed routing.

\begin{table}
\centering
\scriptsize
\begin{tabular}{||p{105pt}p{14pt}p{14pt}p{27pt}p{27pt}||}\hline\hline
\multicolumn{1}{||c}{SFs Elements} & Half Byte & Half Byte & One Byte & Two Bytes \\\hline\cline{2-2}
Active SFs & \multicolumn{1}{|c|}{4 bits}  & & & \\\cline{2-2}\cdashline{3-3}
(Byte Filler) & & \multicolumn{1}{|c|}{4 bits} & & \\\cdashline{3-3}
(Ephemeral Single-hop FID) & & \multicolumn{1}{|c|}{4 bits} & & \\\cline{2-3}
(Intra-region FID) & \multicolumn{2}{|c|}{8 bits} & & \\\cline{2-4}
(Intermediate RID) & \multicolumn{3}{|c|}{16 bits\markarrowtopleft{a1}} & \markarrowbottomright{a1} \\\cline{2-4}
($\cdots$) & \multicolumn{3}{|c|}{$\cdots$\markarrowtopleft{a2}} & \markarrowbottomright{a2} \\\cline{2-4}
Destination RID & \multicolumn{3}{|c|}{16 bits\markarrowtopleft{a3}} & \markarrowbottomright{a3} \\\cline{2-4}\hline
(Active IDs SF Elements) & \multicolumn{1}{|c|}{4 bits}  & & & \\\cline{2-2}\cdashline{3-3}
(Byte Filler) & & \multicolumn{1}{|c|}{4 bits} & & \\\cdashline{3-3}\cline{4-4}
(Packet PID)  & & \multicolumn{2}{|c|}{12 bits}  & \\\cline{3-4}
(Flow FID)  & & \multicolumn{2}{|c|}{12 bits}  & \\\cline{2-4}
Sender NID & \multicolumn{3}{|c|}{16 bits\markarrowtopleft{a4}} & \markarrowbottomright{a4}\\\cline{2-4}
Receiver NID & \multicolumn{3}{|c|}{16 bits\markarrowtopleft{a5}} & \markarrowbottomright{a5} \\\cline{2-4}\hline
(Active QoS SF Elements) & \multicolumn{1}{|c|}{4 bits}  & & & \\\cline{2-2}\cdashline{3-3}
(Byte Filler) & & \multicolumn{1}{|c|}{4 bits} & & \\\cdashline{3-3}
(Single-hop Latency) & & \multicolumn{1}{|c|}{4 bits}  & &\\\cline{2-3}
(Path Latency) & \multicolumn{1}{|c|}{4 bits} & & & \\\cline{2-3}
(Single-hop Loss) & & \multicolumn{1}{|c|}{4 bits}  & &\\\cline{2-3}
(Path Loss) & \multicolumn{1}{|c|}{4 bits} & & & \\\cline{2-3}
(Fission Rate) & & \multicolumn{1}{|c|}{4 bits}  & &\\\cline{2-3}\hline
(Active BRS SF Elements) & \multicolumn{1}{|c|}{4 bits}  & & &\\\cline{2-2}\cdashline{3-3}
(Byte Filler) & & \multicolumn{1}{|c|}{4 bits} & & \\\cline{2-2}\cdashline{3-3}\cline{4-4}
(Source [Backward] RID) & \multicolumn{3}{|c|}{16 bits\markarrowtopleft{a6}} &  \markarrowbottomright{a6}\\\cline{2-4}
($\cdots$) & \multicolumn{3}{|c|}{$\cdots$\markarrowtopleft{a7}} &  \markarrowbottomright{a7}\\\cline{2-4}
($2^{nd}$ Immediate Backward RID) & \multicolumn{3}{|c|}{16 bits\markarrowtopleft{a8}} &  \markarrowbottomright{a8}\\\cline{2-4}
(Immediate Backward RID) & \multicolumn{3}{|c|}{16 bits\markarrowtopleft{a9}} &  \markarrowbottomright{a9}\\\cline{2-4}
\hline\hline
\end{tabular}
\caption{A basic syntax for SmartRegion SFs.}
\label{tab_basic_syntax_SFs}
\end{table}

\subsection{A Basic Header Reservation Syntax for SmartRegion}
\label{sec_suppl_Basic_Header_Reservation_Syntax_SmartRegion}
Here a basic syntax for the SmartRegion SFs is proposed. This syntax is only for the purpose of clarifying the concept, and the syntax will be updated in the future.

The syntax, shown in Table \ref{tab_basic_syntax_SFs}, starts with a {\em Active SFs} field that encodes what SFs are presented in the rest of the header. Byte fillers are occasionally used, depending on presence or absence of a field, to keep the boundaries of fields at the byte level. In the presented syntax, we limit the size of fields corresponding to RIDs and NIDs to 16 bits. However, the size of these IDs could be extended by one or two bytes depending on the size of the network (visualized by the red arrows in the table). Also, as before, fields that are labeled in parentheses are optional.

\begin{figure}[!t]
\centering
\setlength{\tabcolsep}{20pt}
\begin{tabular}{@{}cc@{}}
\fbox{\includegraphics[width=2.0in]{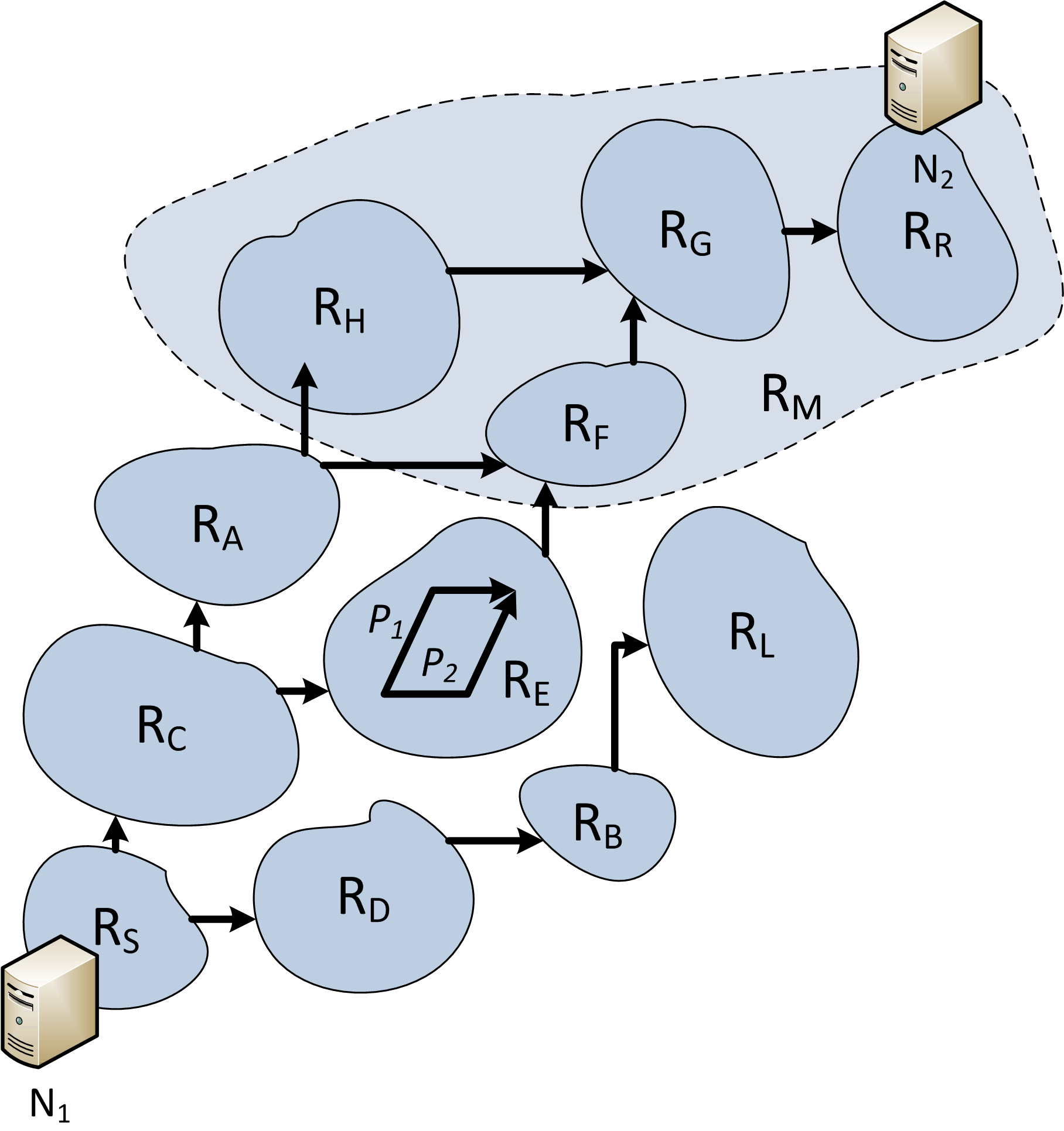}} &
\fbox{\includegraphics[width=2.0in]{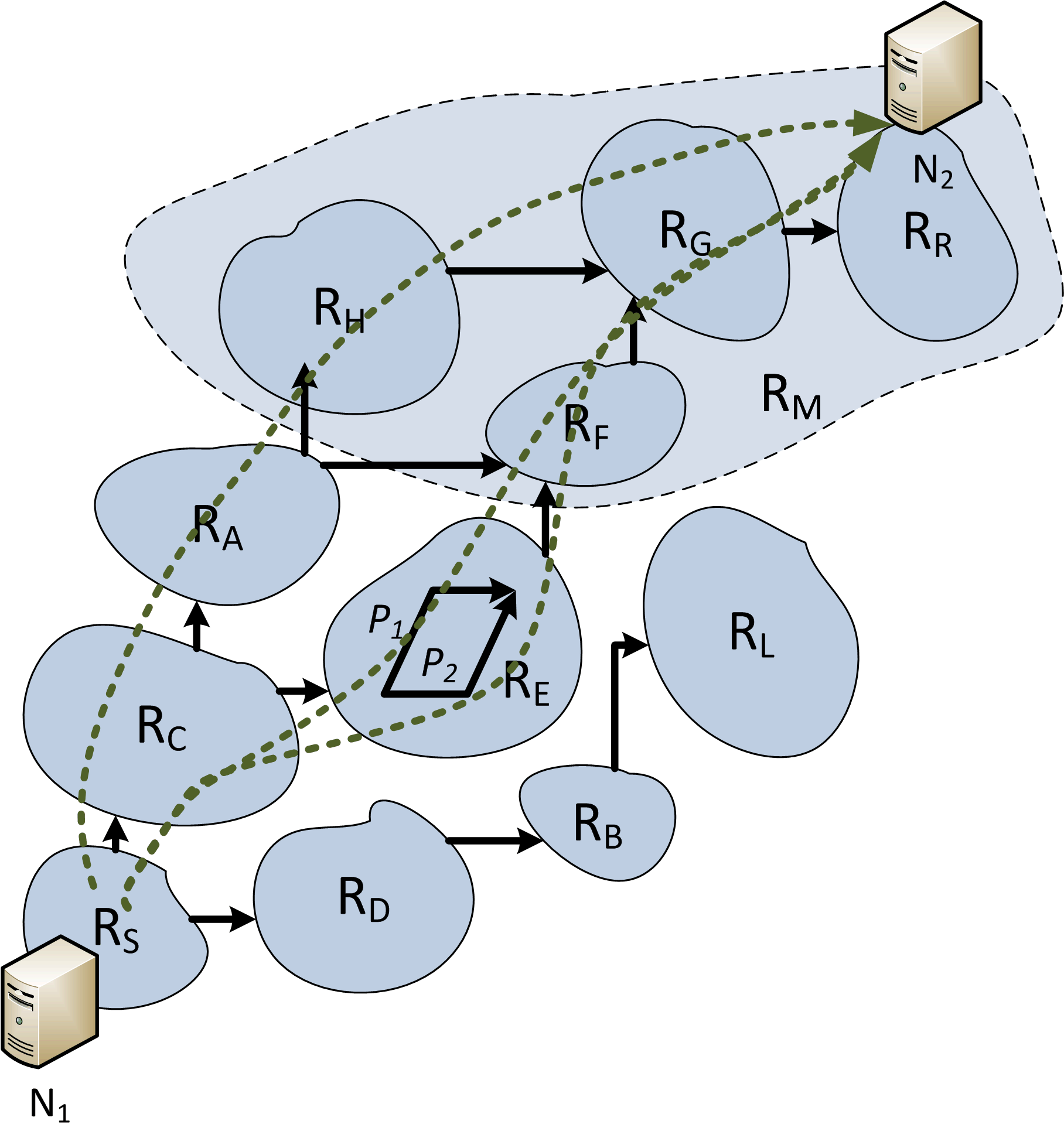}} \\
(a) & (b)
\end{tabular}
\caption{
a) An illustrative region decomposition with only one high-level region $R_M$.
b) A few examples of region paths from a node $N_S$ to a node $N_R$.}
\label{fig_Multipath_Regions_1}
\end{figure}

\subsection{Stationary Routing Protocol}
\label{sec_Stationary_Routing_Protocol}
This mode covers the routing when all the region addresses are up to date. 
In case the region on the top of the Region Stack SF is not the destination region, upon arrival of a packet to a region i) from another region at the same level (neighbor) or ii) from a higher-level region, the receiving switch would generate the possible region-paths for that packet based on packet's Region Stack SuperField. In case the packet belongs to an active flow that have been treated before, the results for other packets of that flow could be recycled. 

In case the region is actually the destination region, the switch would resolve the node-path(s) to the receiver based on the region's connectivity map. The Smart SuperField would be ignored at this level because the delay in processing such data could be higher than the actual `best practice' of the destination region. 

Smartness comes in two forms: The Region Stack SF and The QoS Smart SF. In the first form, a packet could bring its own completely resolved region path with itself that would override all possible decisions of the switches (in terms of region routing, not those related to policies). The second smart capability comes from the QoS Smart SF (as mentioned before); a packet (or a flow) can bring its own specific recipe in terms of QoS, and then it can carry out (possibly fee-bearing) micro-transactions with each region along its journey to the destination region.

Figure \ref{fig_Multipath_Regions_1} provides an illustrative region decomposition and a pair of sender/receiver nodes. Some of possible region paths are shown in Figure \ref{fig_Multipath_Regions_1}(b). In Table \ref{tab_SmartRegion_Header_EX1}, three possible examples of how to fill the Region Stack SF of a packet generated in region $R_S$ with various levels of details. The resolved possible region paths  generated by the intermediate region $R_C$ are shown in Tables \ref{tab_SmartRegion_Header_EX1_resMin} and \ref{tab_SmartRegion_Header_EX1_resMax} with a low or high level of peering-related effort provided by the region $R_C$ for the three cases presented in Table \ref{tab_SmartRegion_Header_EX1}. This clearly shows the flexibility of the proposed approach in allowing different levels of participation by the regions.

\begin{table}
\centering
\scriptsize
\setlength{\tabcolsep}{2pt}
\begin{tabular}{@{}>{\mystrut}ccc@{}}
\imagetop{\begin{tabular}{||l|l||}\hline\hline
\multicolumn{2}{||c||}{a) Region Stack} \\\hline
Index & Region RID \\\hline\hline
1& $R_R$ \\\hline\hline
\end{tabular} 
} &

\imagetop{\begin{tabular}{||l|l||}\hline\hline
\multicolumn{2}{||c||}{b) Region Stack SF} \\\hline
Index & Region RID \\\hline\hline
1& $R_H$ \\\hline
2& $R_G$ \\\hline
3& $R_R$ \\\hline\hline
\end{tabular} 
} &

\imagetop{\begin{tabular}{||l|l||}\hline\hline
\multicolumn{2}{||c||}{c) Fully-resolved Region Stack SF} \\\hline
Index & Region RID \\\hline\hline
1& $R_C$ \\\hline
2& $R_A$ \\\hline
$\cdots$ & $\cdots$ \\\hline
5& $R_R$ \\\hline\hline
\end{tabular} 
} 
\end{tabular} 
\caption{Various examples of possible Region Stack SF for a packet arriving at Region R$_C$ corresponding to Figure \ref{fig_Multipath_Regions_1}.}
\label{tab_SmartRegion_Header_EX1}
\end{table}

\begin{table}
\centering
\scriptsize
\setlength{\tabcolsep}{2pt}
\begin{tabular}{@{}>{\mystrut}ccc@{}}
\imagetop{\begin{tabular}{||l|l||}\hline\hline
\multicolumn{2}{||c||}{a) Region Paths} \\\hline
Path Index & Path \\\hline\hline
1& $\left\{R_A\right\}$ \\\hline
2& $\left\{R_E\right\}$ \\\hline\hline
\end{tabular} 
} &

\imagetop{\begin{tabular}{||l|l||}\hline\hline
\multicolumn{2}{||c||}{b) Region Paths} \\\hline
Path Index & Path \\\hline\hline
1& $\left\{R_A\right\}$ \\\hline\hline
\end{tabular} 
} &

\imagetop{\begin{tabular}{||l|l||}\hline\hline
\multicolumn{2}{||c||}{c) Region Paths} \\\hline
Path Index & Path \\\hline\hline
1& $\left\{R_A\right\}$ \\\hline\hline
\end{tabular} 
} 
\end{tabular} 
\caption{The resolved suggestion to the Region Stack SFs examples of Table \ref{tab_SmartRegion_Header_EX1} in minimal-effort mode.}
\label{tab_SmartRegion_Header_EX1_resMin}
\end{table}

\begin{table}
\centering
\scriptsize
\setlength{\tabcolsep}{2pt}
\begin{tabular}{@{}>{\mystrut}ccc@{}}
\imagetop{\begin{tabular}{||l|l||}\hline\hline
\multicolumn{2}{||c||}{a) Region Paths} \\\hline
Index  & Path \\\hline\hline
1& $\left\{R_A,R_H,R_G,R_R\right\}$ \\\hline
2& $\left\{R_A,R_F,R_G,R_R\right\}$ \\\hline
3& $\left\{R_{E,P_1},R_F,R_G,R_R\right\}$ \\\hline
4& $\left\{R_{E,P_2},R_F,R_G,R_R\right\}$ \\\hline\hline
\end{tabular} 
} &

\imagetop{\begin{tabular}{||l|l||}\hline\hline
\multicolumn{2}{||c||}{b) and c) Region Paths} \\\hline
Index & Path \\\hline\hline
1& $\left\{R_A,R_H,R_G,R_R\right\}$ \\\hline\hline
\end{tabular} 
} 
\end{tabular} 
\caption{The resolved suggestion to the Region Stack SFs examples of Table \ref{tab_SmartRegion_Header_EX1} in maximal-effort mode.}
\label{tab_SmartRegion_Header_EX1_resMax}
\end{table}

\subsubsection{Generating Region-Map}
\label{sec_Generating_Region_Map}
It is based on the region graph. For each `visible' region $R_D$ to a region $R_A$, for example, the `immediate' regions on the possible multiple region paths from $R_A$ to $R_D$ are identified and stored in the region map. The immediate regions would be accompanied with QoS measures along those paths. If an immediate region is associated to several node paths, the QoS information could be aggregated to just a single value for the region, or the region could be multiplicated by the number of its node paths each one associated with its own QoS measure. This would depend on the capacity of the current handling switch. Each one of these instances is denoted $R_{E,P_\omega}$, where $E$ is the name of the region and $\omega$ is the name of a node-path within $R_E$. The process of hosting the region graph and calculating the region map could be provided as a service.

\subsubsection{Partitioning the Network in Regions}
\label{sec_Partitioning_Network_Regions}
Although the network graph can be directly processed, a coordinate-based equivalent representation could help to reduce the amount of calculations required to estimate the QoS metrics of various possible paths (for example, the simple shortest-path metrics). It has been observed that hyperbolic embedding approaches could be used to map network's vertices on a multi-dimensional hyperbolic geometric space \cite{Ajwani2014,Qi2013,Verbeek2014,Adcock2014}. These spaces resemble well the structure of aggregation-based topologies. In future, we will use these representations along with hierarchical clustering techniques to automatically identify the regions that form a region decomposition in its weak form.

\subsection{Multipath Routing in SmartRegion}
\label{sec_suppl_Multipath_Routing_SmartRegion}
It is worth mentioning that multipath routing has been considered before, especially in controller-based solutions. The SmartRegion routing brings two unique multipath features thanks to its region-based approach. However, if these features are not considered in a particular implementation of SmartRegion, a multipath functionality could not be expected:

\subsubsection{Intra-region Multipath Routing} In SmartRegion routing, every region has the autonomy to plan its own way of handling receiving and transiting packets and flows. Although we will discuss various possible strategies for intra-region packet routing and handling in the future, it is straightforward to plan a strategy among switches of a region that makes full benefit of the intra-region connections because of full and transparent visibility available to those switches. This dynamic, real-time optimization of transport resources of a region would be highly {\em hidden} to switches and nodes outside that region. This reduces the complexity of routing to a large degree. 

\subsubsection{Inter-region Multipath Routing} When we look at the network as a graph of interconnected regions, by ignoring internal connectivity and details of each region, routing a packet from a source region $R_S$ to a destination region $R_R$ can be considered as a general routing problem. Considering the lower number of regions compared to the number of nodes/switches, this means that many multipath routing algorithms even with moderate performance could be used with negligible negative impact on the overall performance. The output of such algorithms would be multipath suggestions for a packet (or a flow) at the region level. It is worth noting that the inter-region multipath routing actions are independent from those actions performed by any of the intermediate regions in the form of intra-region multipath routing. However, planning the inter-region routing based on the performance of all the regions considering their intra-region transit performance is suggested and will be considered, especially when OoS constraints are presented.

\subsection{Decentralized Routing Intelligence in SmartRegion}
\label{sec_suppl_Decentralized_Routing_Intelligence_SmartRegion}
The decentralized features of the SmartRegion routing are discussed here. Various form of the routing intelligence could be considered within SmartRegion approach, including:
\begin{enumerate*}[font=\bfseries]
\item Region-Routing Table (at Switch),
\item Region Lookup Service: \begin{enumerate*}[label=\roman*., font=\bfseries]
\item Central for every Region and
\item Multi-Region (Hierarchical),
\end{enumerate*}
\item Region Lookup eXchange (With Neighboring Regions).
\end{enumerate*}
Here we discuss the Region-Routing Table variation in greater details. Full description of all variations will be presented in another work. 

In the Region-Routing Table decentralized routing, every switch is equipped with a region-routing table that in its simplest form it provides: i) the [multiple-valued] next-on-the-path-region for regions and ii) the connected switch(es) of its self region to those next regions. In brief:
\begin{enumerate*}[label=\alph*), font=\bfseries]
\item The region-routing table of a switch could have an optional permanent section that is populated with some preset values associated to very frequent regions of the network. The dynamic part of the tables is gradually populated by adding the information related to the regions on the top of the region stack of packets that arrive at the switch,
\item If the region on the top of the region stack of a packet is unlisted in the region-routing table, the switch pulls the information other switches of the self region. In turn, the switches at the boundary of the self region, would pull the required information from the connected switches of their adjacent regions, \label{list_region_table_step1}
\item An optimal next region is then selected from the region table according to QoS, intra-region and inter-region [load balancing] policies in place,
\item According to the intra-region policy of the self region, the [multi-]path to the connected switch is [retrieved or] planned and then executed, and
\item The region table is updated in a periodic way or if a notification of major change in the network is received.
\end{enumerate*}
Assuming that some of switches have higher level of compute power, a [partial] region map could be hosted in those switches. In this case: 
\begin{enumerate*}[start=6, label=\alph*), font=\bfseries]
\item If the region on the top of the region stack of a packet is unlisted in the region table, the switches calculates the {\em optimal} path to that region from the region map it is hosting. \label{list_region_map_step1}, and 
\item If the step (\ref{list_region_map_step1} fails, other switches are recursively inquired as described in the step (\ref{list_region_table_step1}.
\end{enumerate*}

In summary, in the Region-Routing Table approach, it is assumed that there is a region table that lists the possible next immediate regions if a packet is supposed to routed from a row region $R_\text{A}$ to a column region $R_\text{B}$, where the latter is the region on the top of the region stack of the packet. An element of such a table could have more than one value, which shows the potential of multipath routing at the inter-region level. It is assumed every switch or at least those switches at the boundary of every region host such a table. As mentioned before, there is no need to list all regions of a region decomposition, and listing of a certain number of HCRs would be sufficient assuming that they cover the whole network. A side feature of this flexibility is that the region tables of two regions or even two switches in the same region could be different. In general, considering the low number of regions compared to the total number of switches and nodes, the region tables (and maps) of this approach are finite and small compared to the whole network itself.

\subsection{Dynamic Actions in SmartRegion}
\label{sec_suppl_Dynamic_Actions_SmartRegion}
In this section, we highlight the strategies of SmartRegion routing in the case of failure of a link or a switch. 
First, we would like to mention that we do not want to limit the extent of the proposed routing by choosing a single approach. In other words, we accept various approaches to routing as long as they comply to the region-based notion of SmartRegion. A list of possible approaches has been provided in Section \ref{sec_suppl_Decentralized_Routing_Intelligence_SmartRegion}.  Here, we focus on the Region-Routing Table variation. 
In brief, there are two possible ways to update the region tables and region maps:
\begin{enumerate}
\litem{Explorer Packets} These packets are issued by all switches on a periodic manner, considering the level of congestion in the network, to discover the latest region map of the network. An Explorer Packet is an empty-body packet that is issued to all neighboring regions of a region. Upon receiving such a packet, the receiving switch adds its region to the Backward Region Stack SF, and then return the packet to the issuing region while replicating the packet and forwarding it to all its own neighboring regions. 
\litem{Event Packets} The second mechanism to update the region tables and maps is delivered by Event Packets. These packet are issued by the switches of a region, which has been gone through a change, to inform the other regions. Similar to Explorer Packets, and in general all operations in SmartRegion, hand-over and transit of event packets to neighboring regions is expected to be carried out by the switches of a region upon receiving an event packet either as an act of curtsy (a peering act) or according to their agreements with other regions. 
\end{enumerate}

\subsection{Transitional Routing Protocol}
\label{sec_Transitional_Routing_Protocol}
In the mobile networks and also in the virtual networks, changes in the location of a node are common events. In the period of time that a change in the location is being propagated to update all applications and nodes engaged with the displaced node, handling of the flows and packets requires special attention. We call it transitional routing, and it is described in this section.  When a (destination) node moves to another region, the source node would update the packets of any active flow with the new location's region stack. To recover those packets already in transit, the destination node would issue a propagating message that would ask any potential handling switch to redirect packets of the node's flows to the new region by updating the region stack of the packets. In addition, those packet, which arrive to the old region hosting the destination node, would be forwarded to the new region by the region's switches (and the controller) up to a certain period of time as a gesture of collaboration. Furthermore, the old region would also initiate messages to the source nodes to inform it about the new region again up to a certain period of time, probably longer than that of forwarding gesture.

\section{Discussions and Applications}
\label{sec_Discussions}
The proposed routing can be seen as an enabler to make a packet network cognitive. Here, we provide two examples of possible application beyond simple packet forwarding:

\subsection{QoS Smart SuperField and Net Neutrality}
\label{sec_QoS_Smart_SuperField_Net_Neutrality}
The QoS Smart SuperField of the SmartRegion headers could bring intelligent decision makings for any interested flows in order to take advantage of selecting a region path with better performance. However, this would probably be translated into some form of fee-for-quality agreements and micro-transactions between the sender node (or party) and a (probably access) region. In a generic form, there is no harm of such agreements. However, in the context of the Internet, and especially when paid ISPs and NSPs are involved, an already-payer-for-access receiver party is entitled to some level of immunity that could be seen under the umbrella of the net neutrality and fairness \cite{Kramer2013,Farrahi2014e}. Although we do not consider the case of the Internet in this work, it is worth mentioning that implementing net neutrality does not require abolishing smart approaches, such as those enabled by the QoS Smart SuperField. Instead, it is suggested to enforce fairness by allocating a minimal untradable path and bandwidth that is served based on the multi-tenancy policies \cite{Farrahi2014e}.

\subsection{Next-Region Reservation (NR$^2$) Application}
\label{sec_app_Next_Region_Reservation_NR2}
In this section, we discuss a possible application of the proposed region-based approach in delivering broadband content to nodes on fast moving cars, such as fast train, without requiring constant access along the transportation route. We call this use case the Next-Region Reservation (NR$^2$) process. It is worth mentioning that we assume a 5G-grade intermittent mobile access in this use case. This would require a minimum of 50~Mbps-100~Mbps bandwidth everywhere and 1~Gbps bandwidth indoor. In the current 5G vision based on the evolution of the long term evolution-advanced and its enhancements planned for 2020 \cite{4GAmericas2014,NGMN2015,Ericsson2015,NTTDOCOMOInc2014}.

In a NR$^2$-enabled broadband access, a particular node on a moving car would initiate a NR$^2$ request just before entering a segment of the route without access. In the NR$^2$ request, the immediate network region that the node (along all other riders of the fast train) would enter is provided, and the network would direct all ingress packets and flows associated to that node to the `reserved' region. Assuming that the node's buffer is big enough to survive the biggest non-access segment, the only requirement to achieve a seamless and continuous video experience would be capability to re-buffer in the segments with access. 

In particular, for a fast train at a speed of 300~km/h and a HD video stream encoded using H.265/HEVC @ 2.6 Mbps, a 10~MB video segment file would survive a non-access length of 2.5~km. Considering a 5G-grade access at 50~Mbps, the video segment file would require an access window of 1.6~seconds which is equal to a small cell of an access footprint of 134~m in diameter. This means that only 5\% of the route would be required to provide broadband access. In contrast, if the NR$^2$ process is considered along the 4G-grade access, with a best-practice download rate of 18.4~Mbps, the access time window, the small cell diameter, and the percentage of route to be covered would be 8.2~seconds, 684~m, and 27\% respectively. That means that (NR$^2$, 5G) combination would have require 96\% less power consumption and therefore environmental footprint. It is worth mentioning that without NR$^2$, neither 4G nor 5G could probably provide continuous video experience with the mentioned small diameters mostly because of the latency associated with re-registering and other actions when entering a small cell premises after a non-access segment of the route. 

It is worth mentioning that the action of redirecting the flows to the reserved region could be planned in a smart way in order to prevent requiring a big storage capacity in the small cells. For example, the flows could be delayed or sequenced in such a way that they arrive to the reserved region just-in-time the fast train also passes by.

\begin{figure}[!t]
\centering
\setlength{\tabcolsep}{2pt}
\begin{tabular}{@{}c@{}}
\imagetop{\fbox{\imagewithtext[width=1.11\linewidth]{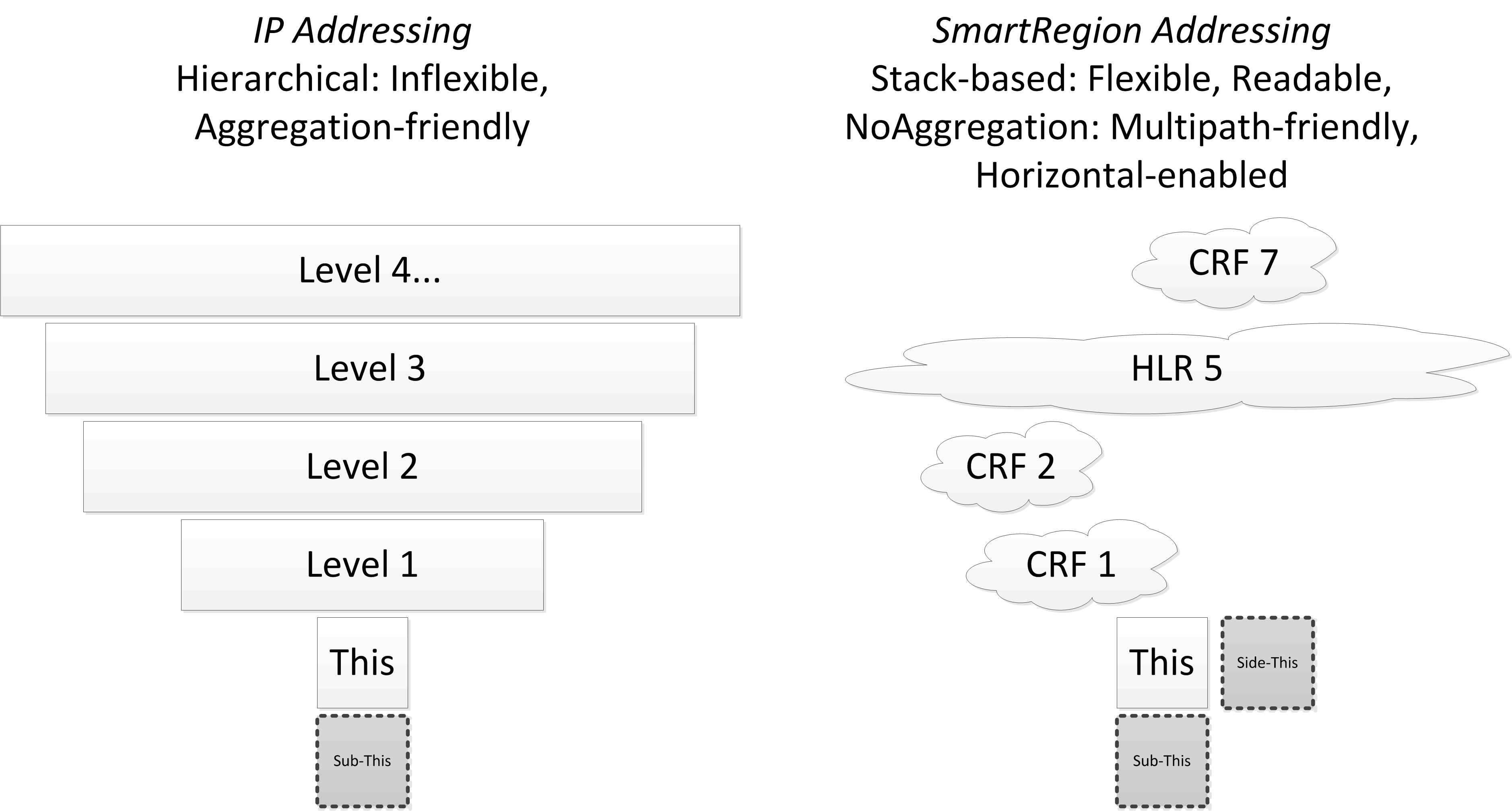}{IP Addressing Hierarchical: Inflexible, Aggregation-friendly. SmartRegion Addressing Stack-based: Flexible, Readable, NoAggregation: Multipath-friendly, Horizontal-enabled. Sub-This. Side-This.}}}
\end{tabular}
\caption{A schematic compression of number-based, aggregation-friendly addressing with the proposed set-based, horizontal-enabled SmartRegion addressing.
}
\label{fig_Region_vs_IP_1}
\end{figure}

\subsection{Application in Addressing}
\label{sec_app_in_addressing}
Although the focus of this work was on using the proposed SmartRegion and Region Path (in the form of the Region Stack) in routing applications, the same concept could be easily generalized in order to be used for the purpose of actual addressing the nodes. This generalization is schematically compared with the more traditional number-based approaches, such as that IP addresses, in Figure \ref{fig_Region_vs_IP_1}. As can be seen, the address of `This' in the SmartRegion addressing has the flexibility of being horizontally expressed. This is highly important for the emerging cases of high-populated networks of devices or {\em things} which naturally scale along horizontal dimension in a heterogeneous manner. In addition, the proposed addressing is inherently multipath-enabled because of the freedom of every individual region to plan the intra-region paths. Moreover, the region-based addressing has the capability to extend `side'-wisely in addition to the traditional `sub'-wise extension planned in IPv6 addressing, for example. 

\begin{figure}[!t]
\centering
\fontsize{10.5}{11.5}\selectfont
\setlength{\tabcolsep}{1pt}
\begin{tabular}{@{}cc@{}}
\imagetop{
\begin{tabular}{||c||c|c|c|c||}\hline\hline
\multirow{2}{*}{Matin Encoding} & \multicolumn{4}{|c||}{Bits 2 and 1} \\\cline{2-5}
 & 00 & 01 & 10 & 11 \\\hline\hline
Bits 5, 4, and 3 & & & & \\\hline\hline
000 & 0 & 1 & 2 & 3 \\\hline
001 & 4 & 5 & 6 & 7 \\\hline
010 & 8 & 9 & A & B \\\hline
011 & C & D & E & F \\\hline
100 & G & H & I & K \\\hline
101 & L & M & N & O \\\hline
110 & P & R & S & T \\\hline
111 & W & X & Y & Z \\\hline\hline
\end{tabular}
} & 
\\\imagetop{\fbox{\imagewithtext[width=0.73\linewidth]{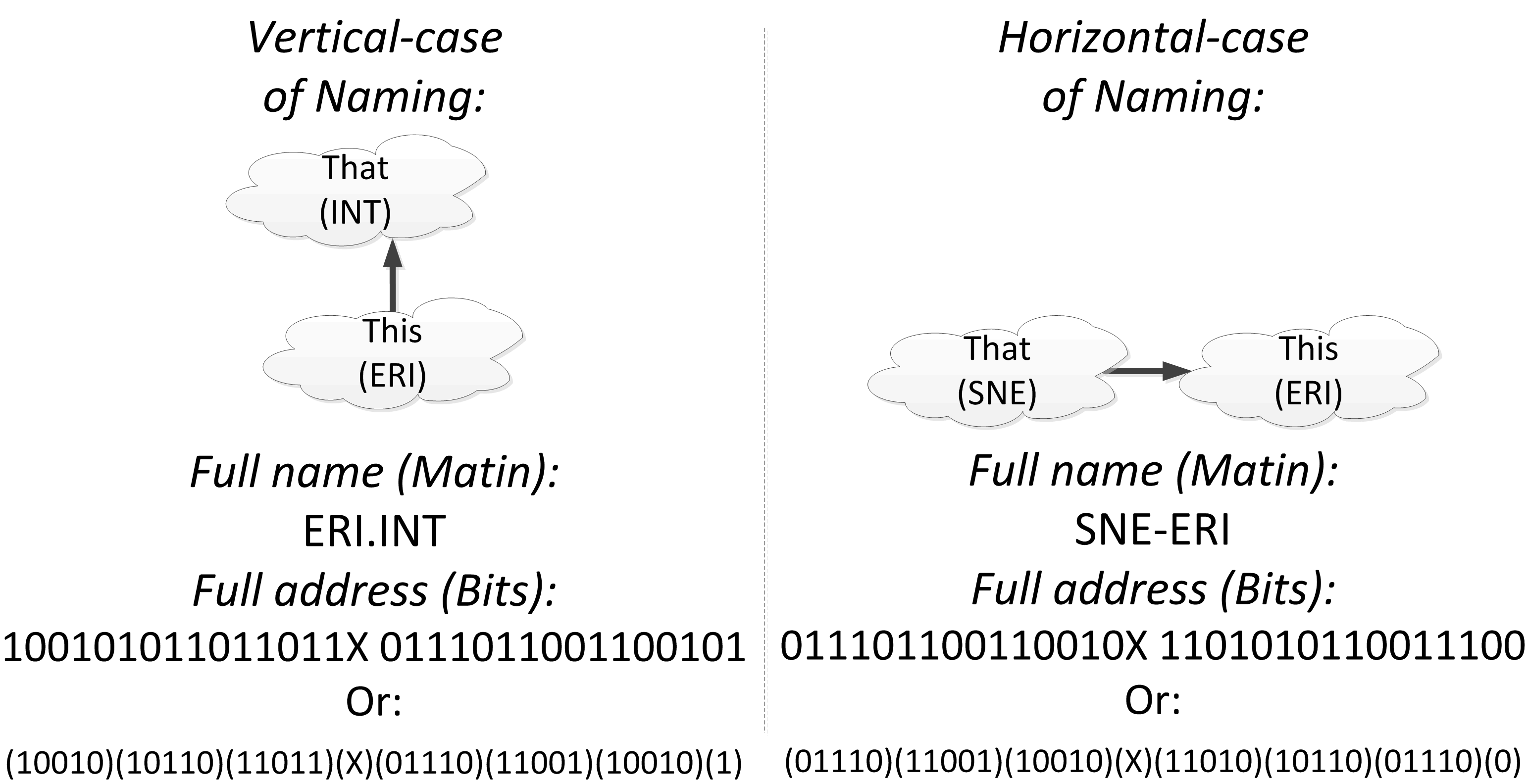}{Vertical-case of Naming:   Horizontal-case of Naming: That (INT) This (ERI) Full name (Matin): ERI.INT SNE-ERI Full address (Bits): 100101011011011X 0111011001100101 Or: (10010)(10110)(11011)(X)(01110)(11001)(10010)(1)}}}\\
(a) & (b)
\end{tabular}
\caption{
a) The 5-bit encoding of the 32 characters of the proposed Matin encoding. 
b) Examples of Matin naming of two-region cases of vertical and horizontal relations.
}
\label{fig_Region_vs_IP_naming_1}
\end{figure}

\subsection{Application in Naming}
\label{sec_app_in_naming}
A step further that region-addressing of Section \ref{sec_app_in_addressing} is using the region-based approach toward a new naming strategy. Almost in all ICT-related activities, naming of the involved entities, ranging from devices, to VMs, among others, has been a challenge. Balancing between readability of the names and also uniqueness is a major barrier in this direction. Here, we proposed to use the `flattened' region stack as framework for readable names. Considering minimal-length units of 16~bits, each unit is encoded as a sequence of 3 characters in a compact {\em Matin}, introduced in Figure \ref{fig_Region_vs_IP_naming_1}(a). The Matin encoding stands for Minimal-Latin and has 32 characters expressed using 5~bits. The last bit of the 16-bit unit is used to indicate whether the next (above) unit is actually vertical (bit=1) or horizontal (bit=0). In the actual name, we translate this bit in `.' or `-' depending if it is 1 or 0. An example of the proposed naming approach is provided in Figure \ref{fig_Region_vs_IP_naming_1}(b). It is worth noting that the order of units when written as the name in the Matin encoding is set in the reverse order of the actual bits of the address in order to make it more human-friendly. The proposed region-based naming is capable to provide
\begin{enumerate*}[font=\bfseries]
\item human-readable,
\item horizontal- and vertical-friendly, and 
\item byte-consistent
\end{enumerate*}
names. Especially, it is worth mentioning that our proposed approach practically integrates both the address and the name of a node in one property in such a way that the same bit sequence would actually express both the address and the name (of a node or any other entity). This would reduce or eliminate many name-related services and their associated overhead and latency. In terms of the capacity, even if we limit the name only to vertical relations, a considerable number of $2^{8\times(16-1)}\simeq 1.3\times 10^{36}$ unique `names' are available in the proposed naming approach assuming a 128-bit address format (compared to the $3.4\times 10^{38}$ unique `addresses' of the IPv6). We will further investigate this naming approach in another work in the future.

\section{Conclusions and future prospects}
\label{sec_conclusion}
A disaggregation-oriented, region-based approach to software defined networks is introduced. Using the notion of full regions, a consistent way to represent a network by its (probably non-unique) regional decomposition is provided. Although the regional decomposition of a network and the network itself are not equal, the region-level representation provides a great potential to decouple routing mechanism between inter- and intra-region levels. This in turn provides possibility to develop high-level and simplified inter-region management of the network while allowing the regions govern their intra-region activities in a semi-autonomous manner. Therefore, the proposed SmartRegion could provide a highly scalable solution that rapidly adapt to the changes in the network which are inevitable considering the mobility and dynamicity of the emerging networks. To implement the SmartRegion routing, an associated header is defined that composed of four SuperFields. The main SuperField is the region stack that smoothly absorbs most of the complexity of the inter-region connectivity, and reduces the associated actions at every intermediate region to only those related to the region on the top of the region stack. Other SuperFields allow the packets and flows to dynamically bring their own QoS along with the capability to track back the packets' route for the purpose of responding to attacks and also higher visibility in the network. In addition to the routing mechanisms, the proposed region-based approach could have various applications in providing seamless broadband services on fast trains, in generalizing addresses beyond the limits of IP addressing, and in integrating and converging address and name in only one identifier. 

Generalization to other forms of routing, such as that of circuit-based networks, will be considered in the future. South-wise translation of SDNs on their underlying physical layer will be considered in order to extend the level of reachability and penetration of the proposed approach. This would also require north-wise visibility and discoverability of the underlying physical layer. All these directions will be exploited toward an extended, multi-layer region decomposition and its associated routing mechanisms.

\iftoggle{noblindflag}{\section*{Acknowledgment}
The authors thank the NSERC of Canada for their financial support under Grant CRDPJ 424371-11 and also under the Canada Research Chair in Sustainable Smart Eco-Cloud.
}

\begingroup
\bibliographystyle{agsm} \bibliography{imagep}
\endgroup

\section*{Appendix}

\setcounter{section}{0}
\renewcommand{\thesection}{Appendix \arabic{section}}

\section{Alternative Definition of Full Regions: REACH}
\label{sec_Alternative_Definition_REACH}
As mentioned in Section \ref{sec_Basic_Definitions}, the DISAGG variation of the SmartRegion approach is not the only possibility. Here, the REACH alternative to definition of the regions in the SmartRegion is introduced. This alternative focuses on maximizing the reachability of nodes. In particular, only the definition of PFRs is modified. An example is provide in Figure \ref{fig_PFR_CFR_Regions_REACH_1}.

\begin{figure}[!t]
\centering
\begin{tabular}{@{}c@{}@{}c@{}}
\imagetop{\fbox{\includegraphics[width=2.6in]{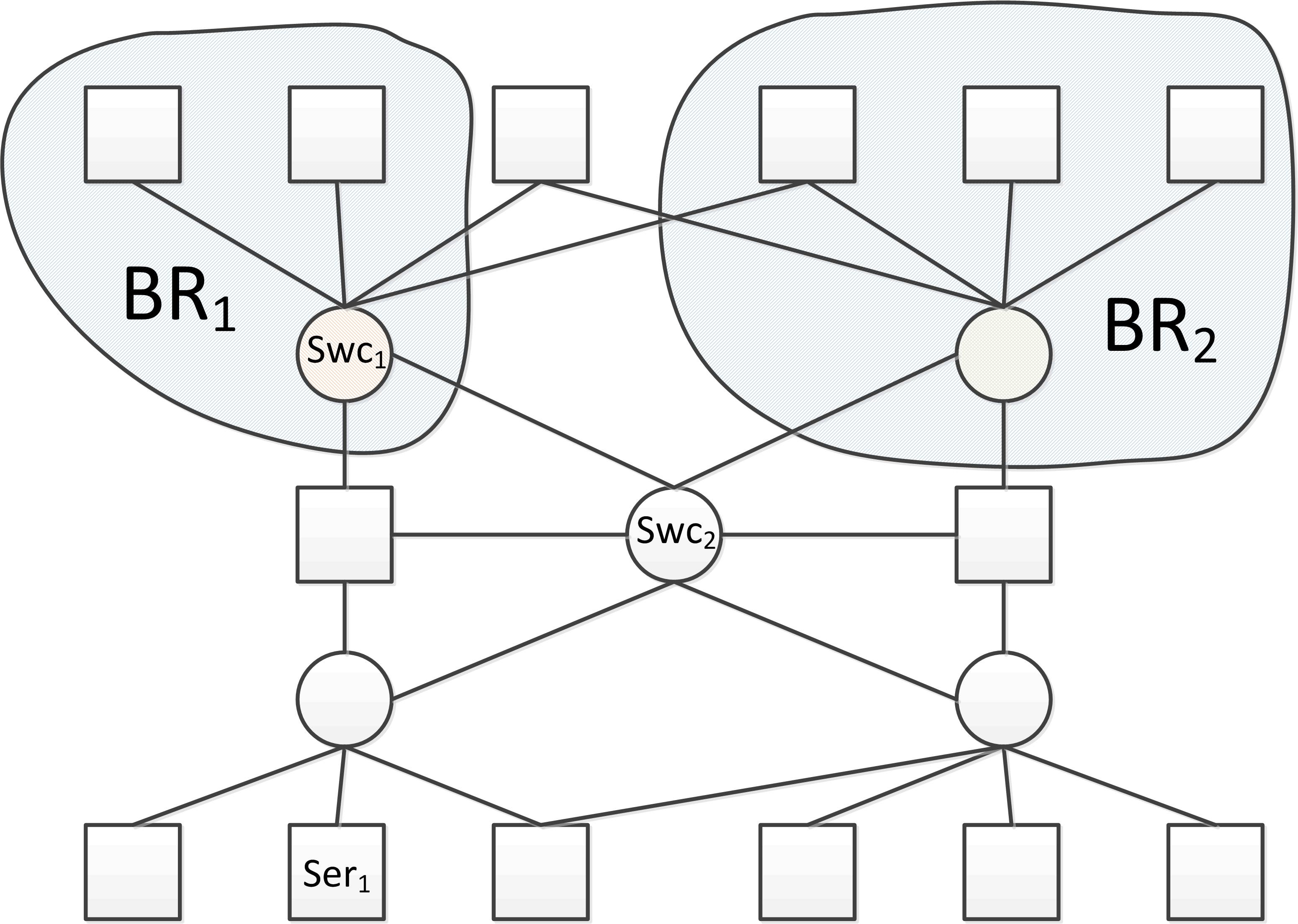}}} &
\imagetop{\fbox{\includegraphics[width=2.6in]{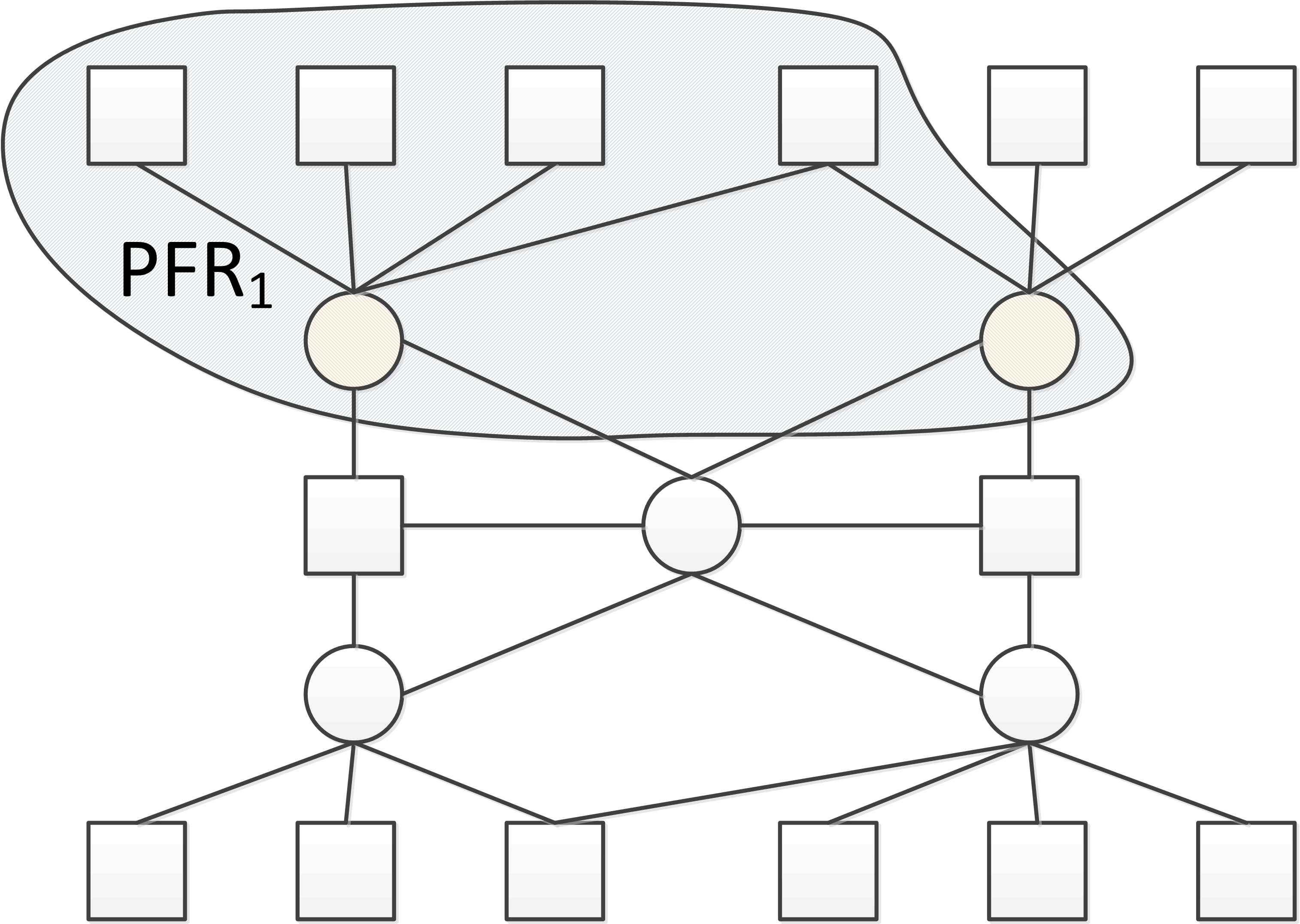}}} \\
(a) & (b) \\
\imagetop{\fbox{\includegraphics[height=1.9in]{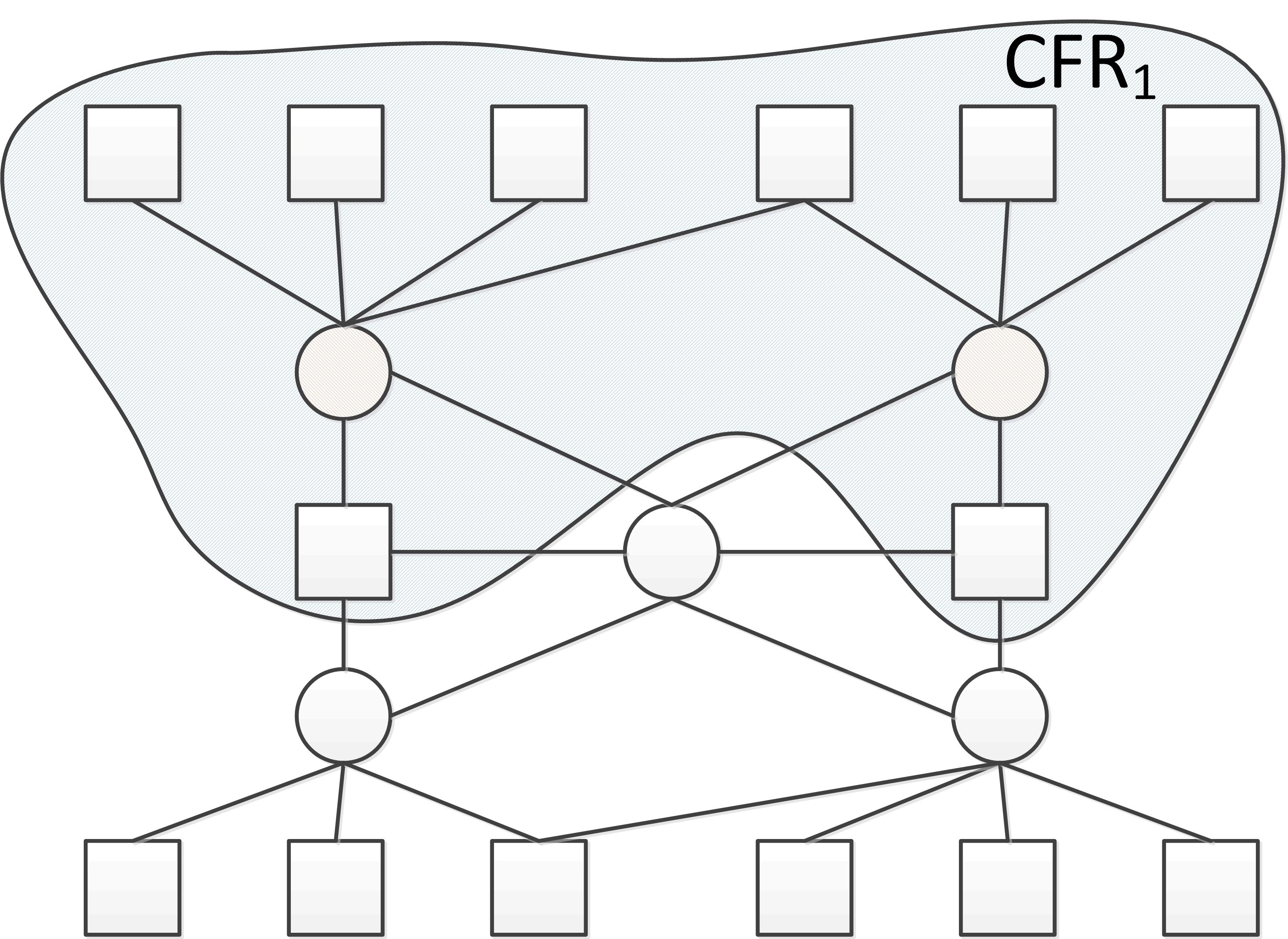}}} &
\imagetop{\fbox{\includegraphics[height=1.9in]{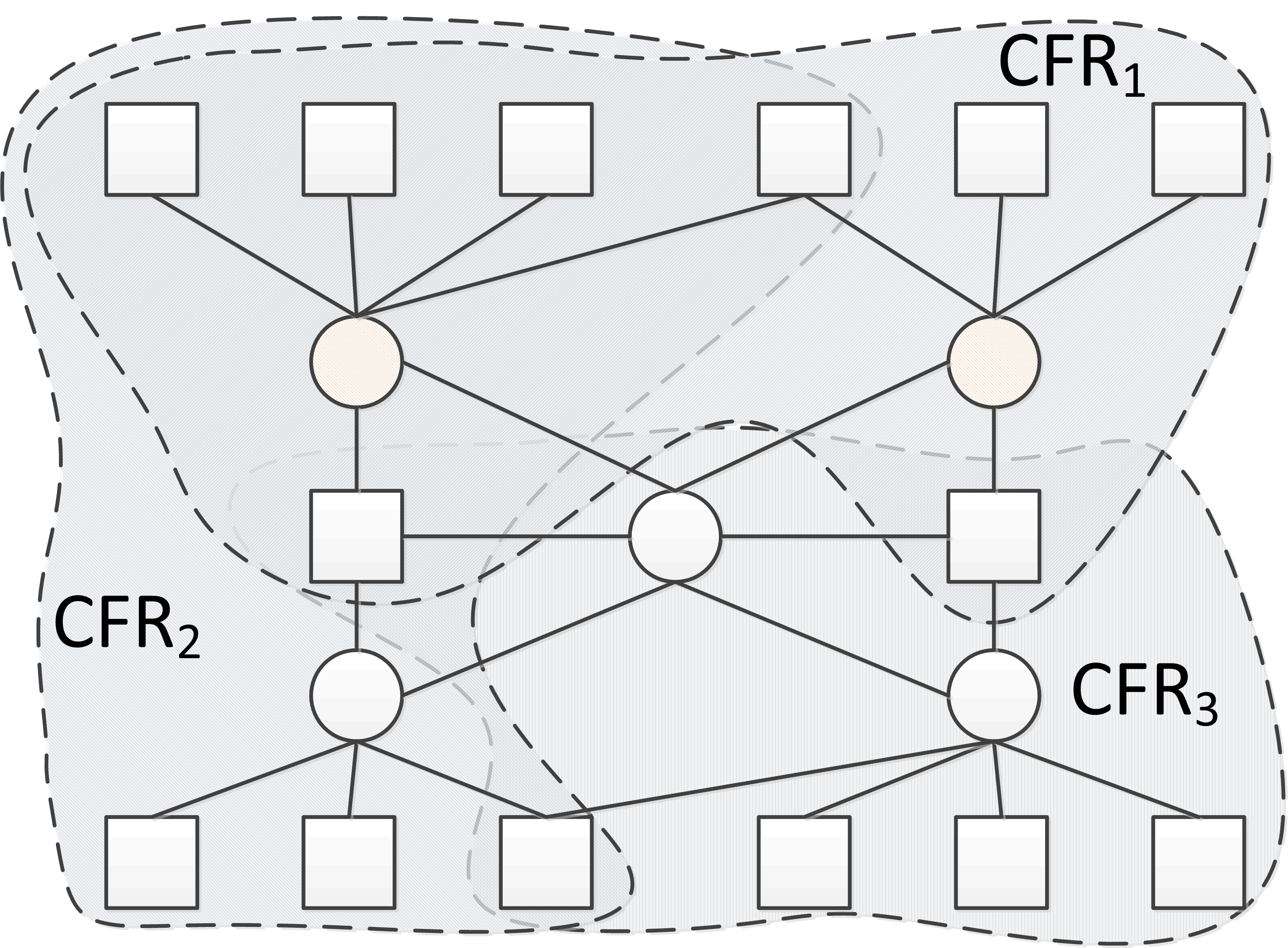}}} \\
(c) & (d) 
\end{tabular}
\caption{An alternative to Figure \ref{fig_PFR_CFR_Regions_1} when the REACH version of the SmartRegion's definition of regions is used. The Basic Regions (BRs) are equivalent. However, Partial-Full Regions (PFRs), Complete-Full Regions (CFRs), and Region Decomposition (RD), provided in (b), (c) and (d), are different.
}
\label{fig_PFR_CFR_Regions_REACH_1}
\end{figure}

\subsection{Partial Full Region} A basic region which includes `all' nodes that are simultaneously connected to all switches of the basic region. Other nodes that do not satisfy this condition are also allowed.

\newpage

\section{Advanced Responding to DDoS}
\label{sec_Advanced_Response_DDoS}
DDoS attacks are complex events and have been long studied from various aspects including detection, mitigation, and responding \cite{KiruthikaDevi2014,Keshariya2010,Chonka2011,Mirkovic2004,Peng2007,Rovniagin2011,Yu2014}. 
\iftoggle{noblindflag}{In \cite{Farrahi2014e}, a}{Recently, a}
proactive approach to response to a Distributed Denial-of-Service (DDoS) attack without completely isolating/blackholing the targeted victim node(s) was proposed. The region-based approaches proposed here, especially the SmartRegion header, have a great potential to realize such responses
\iftoggle{noblindflag}{\cite{Farrahi2015}. 
}{.
}i
In particular, these approaches would open possibilities to implement and execute {\em foris}-wall response mechanisms, which are traditionally initiated blindly by the transport and intermediate networks, in {\em full collaboration} with the victim's access and hosting network. We will study this direction in more details in the future.

\afterpage{\clearpage}

\end{document}